\documentclass[
superscriptaddress,
 amsmath,amssymb,
 aps,
 showpacs,
 prc
]{revtex4-1} 
\usepackage[breaklinks=true,colorlinks=true,linkcolor=blue]{hyperref}
\usepackage{xcolor} 
\usepackage{fdsymbol} 
\usepackage{graphicx}
\usepackage{dcolumn}
\usepackage{comment}
\usepackage[normalem]{ulem} 
\usepackage{xfrac}
\usepackage{bm}
\usepackage[mathlines]{lineno}

\begin{document}

\title{Measurement of the Beam-Normal Single-Spin Asymmetry for \\
Elastic Electron Scattering from $^{12}$C and  $^{27}$Al}%

\author{D.~Androi\'c}
\affiliation{University of Zagreb, Zagreb, HR 10002 Croatia } 
\author{D.S.~Armstrong} 
\email{corresponding author: armd@jlab.org}
\affiliation{William \& Mary, Williamsburg, Virginia 23185 USA}
\author{A.~Asaturyan}
\affiliation{A.~I.~Alikhanyan National Science Laboratory (Yerevan Physics Institute), Yerevan 0036, Armenia}
\author{K. Bartlett}
\affiliation{William \& Mary, Williamsburg, Virginia 23185 USA}
\author{R.S.~Beminiwattha}
\affiliation{Ohio University, Athens, Ohio 45701 USA}
\affiliation{Louisiana Tech University, Ruston, Louisiana 71272 USA}
\author{J.~Benesch}
\affiliation{Thomas Jefferson National Accelerator Facility, Newport News, Virginia 23606 USA}
\author{F.~Benmokhtar}
\affiliation{Duquesne University, Pittburgh, Pennsylvania 15282, USA}
\author{J.~Birchall}
\affiliation{University of Manitoba, Winnipeg, Manitoba R3T2N2 Canada}
\author{R.D.~Carlini}
\affiliation{Thomas Jefferson National Accelerator Facility, Newport News, Virginia 23606 USA}
\author{M. E.~Christy}
\affiliation{Hampton University, Hampton, Virginia 23668 USA}
\author{J.C.~Cornejo}
\affiliation{William \& Mary, Williamsburg, Virginia 23185 USA}
\author{S.~Covrig Dusa}
\affiliation{Thomas Jefferson National Accelerator Facility, Newport News, Virginia 23606 USA}
\author{M.M.~Dalton}
\affiliation{University of Virginia,  Charlottesville, Virginia 22903 USA}
\affiliation{Thomas Jefferson National Accelerator Facility, Newport News, Virginia 23606 USA}
\author{C.A.~Davis}
\affiliation{TRIUMF, Vancouver, British Columbia V6T2A3 Canada}
\author{W.~Deconinck}
\affiliation{William \& Mary, Williamsburg, Virginia 23185 USA}
\author{J.F.~Dowd}
\affiliation{William \& Mary, Williamsburg, Virginia 23185 USA}
\author{J.A.~Dunne}
\affiliation{Mississippi State University,  Mississippi State, Mississippi 39762  USA}
\author{D.~Dutta}
\affiliation{Mississippi State University,  Mississippi State, Mississippi 39762  USA}
\author{W.S.~Duvall}
\affiliation{Virginia Polytechnic Institute \& State University, Blacksburg, Virginia 24061 USA}
\author{M.~Elaasar}
\affiliation{Southern University at New Orleans, New Orleans, Louisiana 70126 USA}
\author{W.R.~Falk}
\altaffiliation{deceased}
\affiliation{University of Manitoba, Winnipeg, Manitoba R3T2N2 Canada}
\author{J.M.~Finn}
\altaffiliation{deceased}
\affiliation{William \& Mary, Williamsburg, Virginia 23185 USA}
\author{T.~Forest}
\affiliation{Idaho State University, Pocatello, Idaho 83209 USA}
\affiliation{Louisiana Tech University, Ruston, Louisiana 71272 USA}
\author{C. Gal}
\affiliation{University of Virginia,  Charlottesville, Virginia 22903 USA}
\author{D.~Gaskell}
\affiliation{Thomas Jefferson National Accelerator Facility, Newport News, Virginia 23606 USA}
\author{M.T.W.~Gericke}
\affiliation{University of Manitoba, Winnipeg, Manitoba R3T2N2 Canada}
\author{V.M.~Gray}
\affiliation{William \& Mary, Williamsburg, Virginia 23185 USA}
\author{F.~Guo}
\affiliation{Massachusetts Institute of Technology,  Cambridge, Massachusetts 02139 USA}
\author{J.R.~Hoskins}
\affiliation{William \& Mary, Williamsburg, Virginia 23185 USA}
\author{D.C.~Jones}
\affiliation{University of Virginia,  Charlottesville, Virginia 22903 USA}
\author{M.~Kargiantoulakis}
\affiliation{University of Virginia,  Charlottesville, Virginia 22903 USA}
\author{P.M.~King}
\affiliation{Ohio University, Athens, Ohio 45701 USA}
\author{E.~Korkmaz}
\affiliation{University of Northern British Columbia, Prince George, British Columbia V2N4Z9 Canada}
\author{S.~Kowalski}
\affiliation{Massachusetts Institute of Technology,  Cambridge, Massachusetts 02139 USA}
\author{J.~Leacock}
\affiliation{Virginia Polytechnic Institute \& State University, Blacksburg, Virginia 24061 USA}
\author{J.P.~Leckey}
\affiliation{William \& Mary, Williamsburg, Virginia 23185 USA}
\author{A.R.~Lee}
\affiliation{Virginia Polytechnic Institute \& State University, Blacksburg, Virginia 24061 USA}
\author{J.H.~Lee}
\affiliation{Ohio University, Athens, Ohio 45701 USA}
\affiliation{William \& Mary, Williamsburg, Virginia 23185 USA}
\author{L.~Lee}
\affiliation{TRIUMF, Vancouver, British Columbia V6T2A3 Canada}
\affiliation{University of Manitoba, Winnipeg, Manitoba R3T2N2 Canada}
\author{S.~MacEwan}
\affiliation{University of Manitoba, Winnipeg, Manitoba R3T2N2 Canada}
\author{D.~Mack}
\affiliation{Thomas Jefferson National Accelerator Facility, Newport News, Virginia 23606 USA}
\author{J.A.~Magee}
\affiliation{William \& Mary, Williamsburg, Virginia 23185 USA}
\author{R.~Mahurin}
\affiliation{University of Manitoba, Winnipeg, Manitoba R3T2N2 Canada}
\author{J.~Mammei}
\affiliation{University of Manitoba, Winnipeg, Manitoba R3T2N2 Canada}
\affiliation{Virginia Polytechnic Institute \& State University, Blacksburg, Virginia 24061 USA}
\author{J.W.~Martin}
\affiliation{University of Winnipeg, Winnipeg, Manitoba R3B2E9 Canada}
\author{M.J.~McHugh}
\affiliation{George Washington University, Washington, D.C. 20052 USA}
\author{D.~Meekins}
\affiliation{Thomas Jefferson National Accelerator Facility, Newport News, Virginia 23606 USA}
\author{K.E.~Mesick}
\affiliation{George Washington University, Washington, D.C. 20052 USA}
\affiliation{Rutgers, the State University of New Jersey, Piscataway, New Jersey 088754 USA}
\author{R.~Michaels}
\affiliation{Thomas Jefferson National Accelerator Facility, Newport News, Virginia 23606 USA}
\author{A.~Mkrtchyan}
\affiliation{A.~I.~Alikhanyan National Science Laboratory (Yerevan Physics Institute), Yerevan 0036, Armenia}
\author{H.~Mkrtchyan}
\affiliation{A.~I.~Alikhanyan National Science Laboratory (Yerevan Physics Institute),
Yerevan 0036, Armenia}
\author{A.~Narayan}
\affiliation{Mississippi State University,  Mississippi State, Mississippi 39762  USA}
\author{L.Z.~Ndukum}
\affiliation{Mississippi State University,  Mississippi State, Mississippi 39762  USA}
\author{V.~Nelyubin}
\affiliation{University of Virginia,  Charlottesville, Virginia 22903 USA}
\author{Nuruzzaman}
\affiliation{Hampton University, Hampton, Virginia 23668 USA}
\affiliation{Mississippi State University,  Mississippi State, Mississippi 39762  USA}
\author{W.T.H van Oers}
\affiliation{TRIUMF, Vancouver, British Columbia V6T2A3 Canada}
\affiliation{University of Manitoba, Winnipeg, Manitoba R3T2N2 Canada}
\author{V.F. Owen} 
\affiliation{William \& Mary, Williamsburg, Virginia 23185 USA}\author{S.A.~Page}
\affiliation{University of Manitoba, Winnipeg, Manitoba R3T2N2 Canada}
\author{J.~Pan}
\affiliation{University of Manitoba, Winnipeg, Manitoba R3T2N2 Canada}
\author{K.D.~Paschke}
\affiliation{University of Virginia,  Charlottesville, Virginia 22903 USA}
\author{S.K.~Phillips}
\affiliation{University of New Hampshire, Durham, New Hampshire 03824 USA}
\author{M.L.~Pitt}
\affiliation{Virginia Polytechnic Institute \& State University, Blacksburg, Virginia 24061 USA}
\author{R.W. Radloff}
\affiliation{Ohio University, Athens, Ohio 45701 USA}
\author{J.F.~Rajotte}
\affiliation{Massachusetts Institute of Technology,  Cambridge, Massachusetts 02139 USA}
\author{W.D.~Ramsay}
\affiliation{TRIUMF, Vancouver, British Columbia V6T2A3 Canada}
\affiliation{University of Manitoba, Winnipeg, Manitoba R3T2N2 Canada}
\author{J.~Roche}
\affiliation{Ohio University, Athens, Ohio 45701 USA}
\author{B.~Sawatzky}
\affiliation{Thomas Jefferson National Accelerator Facility, Newport News, Virginia 23606 USA}
\author{T.~Seva}
\affiliation{University of Zagreb, Zagreb, HR 10002 Croatia } 
\author{M.H.~Shabestari}
\affiliation{Mississippi State University,  Mississippi State, Mississippi 39762  USA}
\author{R.~Silwal}
\affiliation{University of Virginia,  Charlottesville, Virginia 22903 USA}
\author{N.~Simicevic}
\affiliation{Louisiana Tech University, Ruston, Louisiana 71272 USA}
\author{G.R.~Smith}
\email{corresponding author: smithg@jlab.org}
\affiliation{Thomas Jefferson National Accelerator Facility, Newport News, Virginia 23606 USA}
\author{P.~Solvignon}
\altaffiliation{deceased}
\affiliation{Thomas Jefferson National Accelerator Facility, Newport News, Virginia 23606 USA}
\author{D.T.~Spayde}
\affiliation{Hendrix College, Conway, Arkansas 72032 USA}
\author{A.~Subedi}
\affiliation{Mississippi State University,  Mississippi State, Mississippi 39762  USA}
\author{R.~Subedi}
\affiliation{George Washington University, Washington, D.C. 20052 USA}
\author{R.~Suleiman}
\affiliation{Thomas Jefferson National Accelerator Facility, Newport News, Virginia 23606 USA}
\author{V.~Tadevosyan}
\affiliation{A.~I.~Alikhanyan National Science Laboratory (Yerevan Physics Institute),
Yerevan 0036, Armenia}
\author{W.A.~Tobias}
\affiliation{University of Virginia,  Charlottesville, Virginia 22903 USA}
\author{V.~Tvaskis}
\affiliation{University of Winnipeg, Winnipeg, Manitoba R3B2E9 Canada}
\author{B.~Waidyawansa}
\affiliation{Ohio University, Athens, Ohio 45701 USA}
\affiliation{Louisiana Tech University, Ruston, Louisiana 71272 USA}
\author{P.~Wang}
\affiliation{University of Manitoba, Winnipeg, Manitoba R3T2N2 Canada}
\author{S.P.~Wells}
\affiliation{Louisiana Tech University, Ruston, Louisiana 71272 USA}
\author{S.A.~Wood}
\affiliation{Thomas Jefferson National Accelerator Facility, Newport News, Virginia 23606 USA}
\author{P. Zang}
\affiliation{Syracuse University, Syracuse, New York 13244
USA}
\author{S.~Zhamkochyan}
\affiliation{A.~I.~Alikhanyan National Science Laboratory (Yerevan Physics Institute), Yerevan 0036, Armenia}

\collaboration{The Qweak Collaboration}
\date{Draft 2.0: \today}

\begin{abstract} 
We report measurements of the parity-conserving beam-normal single-spin elastic scattering asymmetries $B_n$ on $^{12}$C and $^{27}$Al, obtained with an  electron beam polarized transverse to its momentum direction. These measurements add an additional kinematic point to a series of previous measurements of $B_n$
on $^{12}$C and provide a first measurement on $^{27}$Al. The experiment utilized the Q$_{\rm weak}$ apparatus at Jefferson Lab with a beam energy of 1.158 GeV. The average lab scattering angle for both targets was $7.7^\circ$, and the average $Q^2$ for both targets was 0.02437 GeV$^2$ ($Q=0.1561$ GeV). 
The  asymmetries are $B_n =  -10.68 \pm  0.90 {\rm \:(stat)} \pm 0.57 {\rm \:(syst)}$  ppm for $^{12}$C and
$B_n =  -12.16 \pm  0.58 {\rm \: (stat)} \pm 0.62 {\rm \: (syst)}$  ppm for $^{27}$Al.
The results are consistent with theoretical predictions, and are compared to existing data. When scaled by $Z/A$, the $Q$-dependence of all the far-forward angle ($\theta < 10^\circ$) data from $^{1}$H to $^{27}$Al can be described by the same slope out to $Q\approx 0.35$ GeV. Larger-angle data from other experiments in the same $Q$ range are consistent with a slope about twice as steep.
\end{abstract}

\maketitle

\section{\label{sec:introduction}Introduction}

Electron scattering has a long history as a powerful technique for probing hadron and nuclear structure~\cite{Walecka:2001gs}. In the case of parity-violating electron scattering, it has also been used to test the electroweak sector of the standard model, and thereby to search for new physics. As the precision of such experiments has improved, it has become necessary when analyzing the data to go beyond the single boson (photon or $Z$) exchange approximation, and include higher-order terms. Such terms include two-boson exchange corrections, e.g. $\gamma \gamma$ and $\gamma Z$ diagrams. The former is understood to be of critical importance for measurements of the proton's electric form factor $G_E^p$~\cite{Afanasev:2017gsk}. The apparent inconsistency between the form factor at high four-momentum transfer $Q^2$ as extracted using the Rosenbluth separation technique and that obtained from recoil polarization measurements appears to be at least partially explained by the greater contribution of $\gamma \gamma$ exchange to the Rosenbluth analysis~\cite{Guichon:2003qm,*Blunden:2003sp}. As a second example, the $\gamma Z$ box diagram~\cite{Gorchtein:2008px,*Sibirtsev:2010zg,*Rislow:2010vi,*Rislow:2013vta,*Gorchtein:2011mz,*Blunden:2011rd,*Hall:2013hta,*Hall:2015loa} 
provides a numerically significant contribution to precision measurements of $A_{\rm PV}$, the parity-violating asymmetry  in the scattering of longitudinally-polarized electrons from unpolarized protons, such as in the recent Q$_{\rm weak}$ experiment~~\cite{Androic:2018kni,Androic:2013rhu} and the upcoming P2 experiment~\cite{Becker:2018gzk}. Similar multiboson exchange effects, such as $\gamma W$ and $W Z$ box diagrams, are relevant for precision measurements of other electroweak processes, such as super-allowed nuclear beta decay~\cite{PhysRevLett.121.241804}.
In addition to ``hard" two-boson exchange, higher-order electromagnetic effects due to the 
strong electric field of the nucleus (``Coulomb distortions") may also need to be accounted for when scattering electrons from nuclei with high atomic number~\cite{Article:Aste2005}.

One observable in electron scattering that directly probes two-photon exchange (TPE) is the beam-normal single-spin asymmetry (BNSSA), $B_n$ (or $A_y$~\cite{osti_4726823}). This is a parity-conserving asymmetry, which arises in the elastic scattering of electrons polarized normal to the scattering plane when scattering from an unpolarized target. It is identically zero for pure one-photon exchange, due to time-reversal invariance, and it is generated by the interference between single photon and two photon exchange amplitudes~\cite{DeRujula:1972te}. $B_n$ gives direct access to the imaginary (absorptive) part of the TPE amplitude. $B_n$ is defined as 

\vspace*{-0.3cm}
\begin{equation}
\label{eq:normal_spin_formula}
B_n = \frac{\sigma^\uparrow -  \; \sigma^\downarrow}{\sigma^\uparrow +  \; \sigma^\downarrow} = \frac{2 \; \mathcal{I}m ( \mathcal{M}_{\gamma \gamma} \mathcal{M}_{\gamma}^* )}{|\mathcal{M}_{\gamma}| ^2},
\end{equation}
\noindent
where 
$\sigma^\uparrow (\sigma^\downarrow)$ denotes the scattering cross section for electrons with spin parallel (anti-parallel) to a vector $\hat{n}$ perpendicular to the scattering plane. Here $\hat{n}=(\vec{k}\times\vec{k^{\prime}})/(|\vec{k}\times\vec{k^{\prime}}|)$ with $\vec{k}(\vec{k^{\prime}})$ being the momentum of the incoming (outgoing) electron.  $\mathcal{M}_{\gamma}$ and $\mathcal{M}_{\gamma \gamma}$ are the amplitudes for one- and two-photon exchange. For high beam energies ($> 100$ MeV) this asymmetry was first observed 
over 20 years ago in the SAMPLE parity-violating electron-scattering experiment~\cite{Wells:2000rx}, where it was referred to as the ``vector analyzing power," even though by convention~\cite{osti_4726823} that terminology is meant to refer to observables with a vector polarized target. It is also sometimes referred to as the ``transverse asymmetry.'' At much lower energies ({\em i.e.} a few MeV) this asymmetry is known as the Mott asymmetry, and is used in electron beam polarimetry~\cite{Grames:2020asy}.

The beam-normal single spin asymmetry $B_n$ depends on the imaginary part of the two photon exchange amplitude. In contrast, the effect of TPE on reaction cross sections, which is of relevance for comparison of $e^+$ and $e^-$ cross sections \cite{Rachek:2014fam,*Adikaram:2014ykv,*Henderson:2016dea} and for the Rosenbluth determinations of proton form factors~\cite{Guichon:2003qm,*Blunden:2003sp}, depends on the real part of the amplitude. In principle, the real and imaginary parts of the amplitude can be connected via dispersion relations, however this would require data over a broad kinematic range. Nevertheless, measurements of $B_n$ provide a useful benchmark for theoretical models of TPE effects.

Theoretical calculations of TPE needed in order to predict $B_n$ require a model of the doubly-virtual Compton scattering amplitude over a broad range of kinematics, including an inclusive account of intermediate hadronic states, and are therefore challenging. The contribution from these intermediate states  usually dominates the asymmetry, due to the logarithmic enhancement which arises when one of the exchanged hard photons is collinear with the parent electron, as initially recognized by Afanesev and Merenkov~\cite{Afanasev:2004pu}. Several different calculational techniques have been applied  to describe $B_n$  for electron-nucleon scattering. One approach models the intermediate hadronic state in the resonance region via a parameterization of electroabsorption  amplitudes~\cite{Pasquini:2004pv,Tomalak:2016vbf}. In these calculations the hadronic intermediate amplitudes are limited to $\pi N$ states, and the model should apply for all scattering angles. The second approach~\cite{Afanasev:2004pu,Gorchtein:2004ac,Gorchtein:2005yz,PhysRevC.73.035213} uses the optical theorem to relate the doubly-virtual Compton amplitude to the virtual photoabsorption cross section, which therefore encompasses all intermediate states, but is only strictly valid in the forward-angle limit. Heavy baryon chiral perturbation theory has also been used to calculate $B_n$~\cite{Diaconescu:2004aa}, however, this approach is expected only to be applicable for low energy beams.

These models have been confronted with experimental $B_n$ results for the proton and the neutron (deduced from quasielastic scattering on the deuteron), at various $Q^2$, beam energies, and at both forward and backward scattering angles ~\cite{Wells:2000rx,Armstrong:2007vm,Androic:2011rh,Maas:2004pd,Gou:2020viq,Rios:2017vsw,Androic:2020rkw}. The comparison of the data with the
models clearly demonstrates the importance of the inelastic intermediate states in the kinematic ranges that have been studied (see for example~\cite{Armstrong:2007vm,Maas:2004pd,Androic:2011rh,Androic:2020rkw,Rios:2017vsw}), although disagreements between the data and the available models are as large as a factor of two in some cases~\cite{Gou:2020viq,Wells:2000rx}. At very forward angles, there is impressive agreement between the optical model calculations and data,
see~\cite{Androic:2020rkw} and references therein.

The experimentally-measured asymmetry at a given azimuthal scattering angle $\phi$ depends on $B_n$
as \[A_{\rm exp}(\phi) \approx B_n \vec{P}\cdot\hat{n}, \]
 where $\vec{P}$ is the electron polarization vector. 
For beam energies around 1 GeV, elastic asymmetries of order $B_n \approx 10^{-5}$ are expected~\cite{Afanasev:2004pu}. Thus, experiments studying $B_n$ have the challenge of controlling uncertainties below the part-per-million (ppm) level.

The sustained progress in precision measurements of $A_{\rm PV}$ in parity-violating electron scattering over several decades~\cite{Souder:2015mlu,Armstrong:2012bi,Carlini2019} provides both an opportunity and an additional motivation for studying $B_n$. These experiments have demonstrated the ability to control the total uncertainty to well beyond the required level --- the most precise such measurement to date, Q$_{\rm weak}$, achieved a total uncertainty of 0.0093 ppm~\cite{Androic:2018kni}.
In parity-violating asymmetry measurements, which rely on a longitudinally-polarized electron beam, any small transverse components to the beam polarization, combined with non-zero values of $B_n$, may lead to large (on the scale of the desired precision) azimuthally-varying asymmetries, and therefore become important systematic corrections to control. Thus, determinations of $B_n$ for the appropriate kinematics and target represent important ancillary measurements for the parity-violation experiments. Indeed, most of our present experimental information on $B_n$ comes from such measurements.

A related observable to the BNSSA, which also probes TPE, is the target-normal single-spin asymmetry. This can be measured with an unpolarized beam and a target polarized normal to the scattering plane. Here the asymmetries are predicted to typically be much larger than $B_n$ at similar kinematics, {\em  i.e.} of order $10^{-3}$. Only one such measurement has been reported to date, on $^3$He (in quasielastic kinematics to extract the neutron asymmetry), by the Jefferson Lab Hall A Collaboration~\cite{PhysRevLett.115.172502}. The highest-$Q^2$ result was in good agreement with a partonic calculation~\cite{PhysRevLett.93.122301} of TPE.

\section{\label{sec:theory}BNSSA on $A>1$ nuclei}

The situation for $B_n$ for complex nuclei ($A>1$) is less well developed. On the experimental side, the first nuclear measurements were reported for $^4$He, $^{12}$C, and $^{208}$Pb at very forward angle ($\approx 6^{\circ}$) and energies of 1--3 GeV by the HAPPEX/PREX collaborations~\cite{Abrahamyan:2012cg}. More recently, the A1 collaboration at Mainz measured $B_n$ for $^{12}$C at a beam energy of 570 MeV and moderately forward angles ($15^{\circ}$-- $26^{\circ}$), over a range of $Q^2$ (0.023 -- 0.049 GeV$^2$) ~\cite{Esser:2018vdp}.  The same collaboration has also reported measurements for $^{28}$Si and $^{90}$Zr at the same beam energy, at  $Q^2 \approx 0.04$ GeV$^2$~\cite{Esser:2020vjb}.

On the theoretical side, only two approaches have been applied. Cooper and Horowitz~\cite{Cooper:2005sk} addressed the Coulomb distortion effect in a calculation for $^4$He and $^{208}$Pb, using an approach that applies to all orders in photon exchange (not just TPE), by solving the Dirac equation numerically. However, they had to neglect the effect of inelastic hadronic intermediate states. Perhaps for this reason, their calculation did not reproduce the data for either nucleus~\cite{Abrahamyan:2012cg}.

The other approach is the optical theorem approach discussed earlier, which has been extended to complex nuclei by Afanasev and Merenkov~\cite{Afanasev:2004pu} and  by Gorchtein and Horowitz~\cite{Gorchtein:2008dy}.  These calculations work in the forward angle and low-$Q^2$ limit. In that limit the virtual photoabsorption cross section on the nucleon $\sigma_{\gamma^*N}(W,Q^2)$ can be approximated by the real photoabsorption cross section $\sigma_{\gamma N}(W)$ ($W$ is the invariant mass of the intermediate hadronic system). It was shown~\cite{Gorchtein:2005za} that this leaves small corrections of order $Q^2/E_e^2$ where $E_e$ is the electron beam energy. Gorchtein and Horowitz extend this to complex nuclei by noting that the photoabsorption cross section has been measured for a range of nuclei, and was found to be well-reproduced by the nucleon cross section $\sigma_{\gamma N}(W)$, scaled by the mass number $A$~\cite{Bianchi:1995vb}. Thus they use $A\sigma_{\gamma N}(W)$ to represent the nuclear photoabsorption cross section. 

The optical model only rigorously applies in the exact forward angle limit. This approach requires additional input 
for kinematics beyond the forward limit. One needs to account for the $Q^2$ dependence of the Compton scattering amplitude for the nucleus. While the $Q^2$ dependence for the Compton cross section has been measured for the proton~\cite{Bauer:1977iq} and for $^4$He~\cite{Aleksanian:1986hb}, it is not available for other nuclei. It is plausible that this dependence should 
fall more steeply
for nuclei than for the nucleon, in analogy to the observation that the elastic charge form factors are steeper for complex nuclei than for the proton. Consequently, Gorchtein and Horowitz adopt the ansatz that the Compton form factor $F_{\rm Compton}(Q^2)$ for nuclei is approximated by 
\begin{equation}
    F_{\rm Compton}(Q^2) \approx F_{\rm ch}(Q^2) e^{-\frac{B_c}{2}Q^2}, 
\end{equation}
where $F_{\rm ch}$ is the charge form factor of the 
given nucleus, and $B_c$, called the Compton slope parameter, is taken as 8 GeV$^{-2}$. 

This model was observed~\cite{Abrahamyan:2012cg} to predict a simple approximate scaling, at low $Q^2$ and forward angles, for the beam-normal single-spin asymmetry $B_n$ for a given nucleus  of  
\begin{equation}
  B_n \approx \widehat{B}_n \frac{A}{Z}\sqrt{Q^2}  . \label{eq:scaling}  
\end{equation}
Here $A$ and $Z$ are the mass number and atomic number of the element, and $\widehat{B}_n$ is a constant. The model also has the feature that at fixed $Q^2$, $B_n$ is almost independent of the beam energy~\cite{Gorchtein:2008dy,Afanasev:2004pu}.

 With one exception, the Gorchtein and Horowitz model was found to work quite well describing the available data~\cite{Abrahamyan:2012cg,Esser:2018vdp,Esser:2020vjb}, if one assumes an uncertainty in the Compton slope parameter $B_c$ of $\pm 20$\%. The exception is  $^{208}$Pb, where, at the experimental kinematics, the model predicts $B_n \approx -8$~ppm while the measurement~\cite{Abrahamyan:2012cg} yielded $B_n = 0.28  \pm 0.25$ ppm. The cause for this discrepancy is not yet understood. Additional data for $B_n$ on nuclei may shed light on this anomaly. Results are expected for $^{40}$Ca and $^{48}$Ca, as well as new results for $^{12}$C and $^{208}$Pb, from the PREX-2 and CREX experiments ~\cite{McNulty:2020,Richards:2020}.
 
The present measurement extends the data set on $B_n$ in $A>1$ nuclei by providing a forward-angle datum for $^{12}$C  and the first measurement on $^{27}$Al, both at a similar scattering  angle as for the HAPPEX/PREX datum~\cite{Abrahamyan:2012cg} but at a larger $Q^2$.
We note that the $^{27}$Al case represents the first measurement of $B_n$ on a non spin-zero complex nucleus. 
Later in Sec.~\ref{sec:results} the nuclear dependence of $B_n$ will be examined for the three Q$_{weak}$ data on $^1$H~\cite{Androic:2018kni}, $^{12}$C, and $^{27}$Al, all of which were acquired at very similar kinematics ($E$, $\theta$, and $Q$).

\section{\label{sec:experiment}Experiment}

\begin{figure}[!htbp]
\includegraphics[width=1.0\columnwidth]{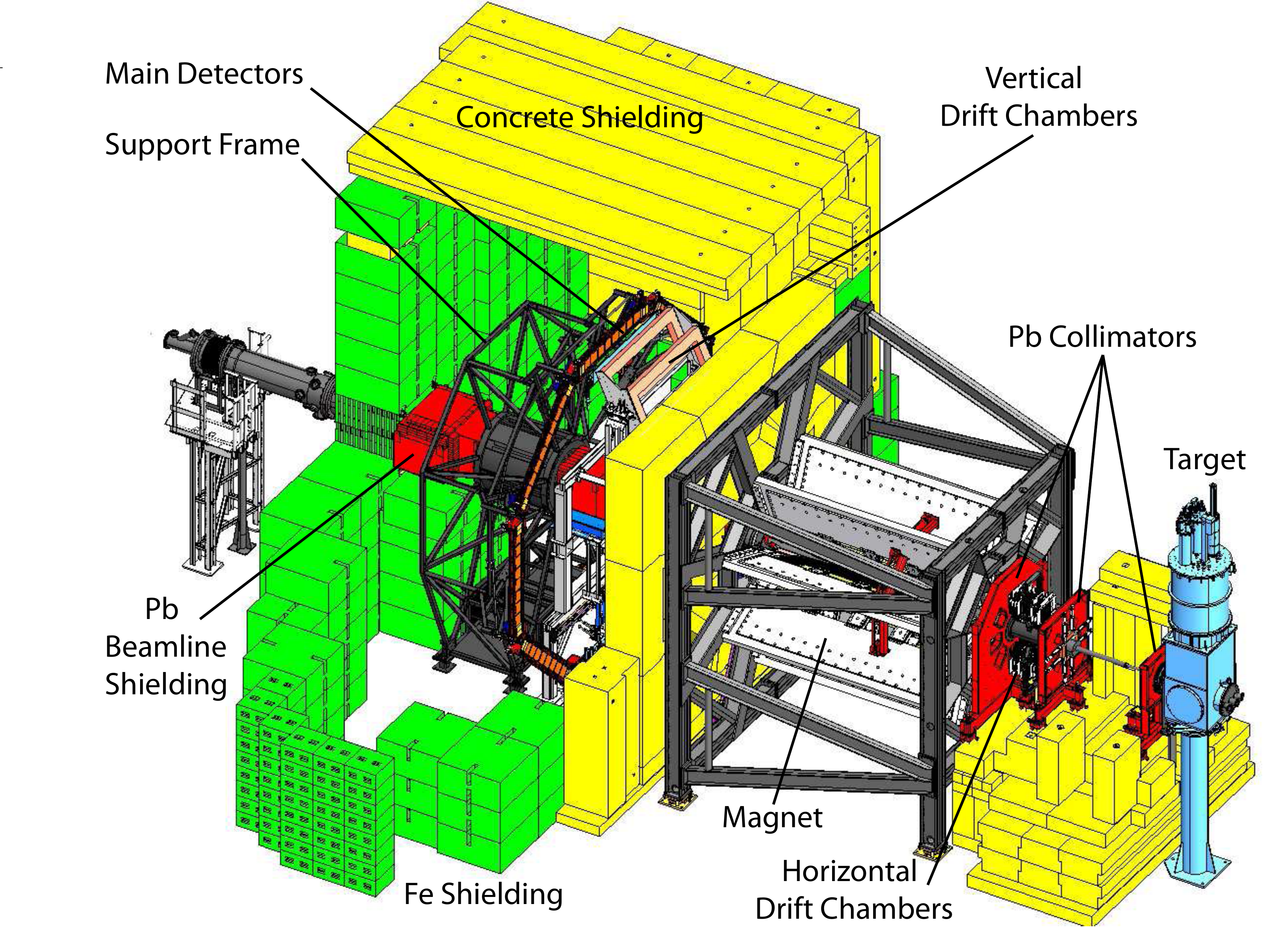}
\caption{
CAD view of the experimental apparatus, which reveals more otherwise hidden details than a photograph can. The beam was incident from the
right. The key elements include the target scattering chamber (cyan), a
triple collimator system (red), a resistive eight-fold symmetric toroidal magnetic spectrometer (grey), and eight Cherenkov detectors (narrow dark orange  octagonal ring in an azimuthally-symmetric array around the beamline at a radius of 3.4 m).  Two tracking drift chambers are illustrated just upstream of the eight Cherenkov detectors; the latter are arrayed about the beamline 12.3 m downstream of the target.  For clarity, only some portions of the extensive
steel and concrete shielding are shown. 
\label{fig:NIM_CAD}
}
\end{figure}

The experiment was performed with the Q$_{\rm weak}$ apparatus, which has been described in detail in~\cite{ALLISON2015105} as well as in the context of the proton's weak charge measurement in~\cite{Androic:2018kni}. Here only a short description of the apparatus (depicted in Fig.~\ref{fig:NIM_CAD}) will be provided.

The 1.158 GeV polarized electron beam was produced by the CEBAF accelerator at the Thomas Jefferson National Accelerator Facility (JLab) and delivered to the Q$_{\rm weak}$ apparatus in experimental Hall C. The  JLab polarized source  directed linearly polarized laser light through a Pockels cell capable of reversing the helicity of the laser light 960 times a second. The helicity  could also be reversed on a slower time scale (typically every 8 h) by inserting a half-wave plate (the IHWP) just before the Pockels cell. The circularly polarized light emerging from the Pockels cell was directed at a photocathode, where the helicity  was transferred to the ejected electrons, which were then accelerated electrostatically. The spin of the nominally longitudinally-polarized electrons was then  rotated to the transverse direction in the injector using a double Wien filter~\cite{Adderley:2011ri}.   The spin direction of the electrons was 
selected  from one of two pseudo-randomly chosen  $\uparrow \downarrow \downarrow \uparrow$ or $\downarrow \uparrow \uparrow \downarrow$ quartet patterns generated at 240 Hz. Here  $\uparrow$ represents the standard spin orientation (spin up or to beam right) and $\downarrow$ represents a 180$^{\circ}$ rotation in the corresponding plane. 

After acceleration through the first of the 5 
passes of the JLab  recirculating linac available at the time, the beam was extracted into the Hall C arc where its momentum could be measured. After that the beam passed through a transport section with beamline instrumentation consisting of beam position monitors (BPMs) and harps~\cite{Yan:1995nc}, beam charge monitors, 
a Compton polarimeter and associated chicane~\cite{Magee:2016xqx,Narayan:2015aua}, a M{\o }ller polarimeter~\cite{Hauger:1999iv}, and air-core raster magnets to spread the nominally 100 $\mu$m (rms) diameter beam across the face of the target in a rectangular 4x4 mm$^2$ pattern. 

Although the systematic uncertainty in the determination of the beam charge normalization was one of the largest relative contributions to the total uncertainty in the very precise parity-violating weak charge measurement on hydrogen~\cite{Androic:2018kni}, it is negligible in the context of the parity-conserving results reported here.

False asymmetries from spin-correlated beam position, angle, and energy changes were largely canceled by a regression algorithm combining beam monitor and scattered electron detector information as described in Sec.~\ref{sec:regression} below. Additional protection against other higher order sources of false asymmetry were largely cancelled by the periodic insertion of the IHWP.
Beam polarization results are discussed in Sec.~\ref{sec:Aexp_corrections}.

The two targets used in this measurement were positioned along the beamline at the same location as the downstream window of the 34-cm-long liquid hydrogen target cell used in the weak charge measurement~\cite{Androic:2018kni}. The relevant kinematics and acceptance parameters discussed below in Sec.~\ref{sec:Aexp_corrections} and ~\ref{sec:results} were all calculated for this position, and the changes were minor relative to those determined for the liquid hydrogen target.
The aluminum target was a 1.032 g/cm$^2$
(3.68 mm thick) 7075-T651 alloy target (4.60\% of $X_0$). This high-strength alloy was fabricated from the same block of material used for the entrance and exit windows of the liquid hydrogen target, so that measurements made on the aluminum target could be used for background
determinations in the weak charge measurement~\cite{Androic:2018kni,Androic:2013rhu}.
The elemental contributions to the alloy were determined by a commercial assay using optical emission spectroscopy~\cite{ATS}.
The carbon target was 99.95\% pure graphite $^{12}$C with an areal density of 0.7018 g/cm$^2$ (3.17 mm thick, or 1.64\% of $X_0$). 

Three Pb collimators centered along the beamline each contained  eight 
openings arrayed symmetrically about the beam axis.
The collimators limited scattering (polar) angles to the range between about $6^\circ < \theta < 11^\circ$, but left open 49\% of the $2\pi$ radians in the azimuthal ($\phi$) direction.  The first collimator 0.5 m  downstream of the target also contained a water cooled W-Cu beam collimator which mitigated small-angle ($\theta > 0.88^\circ$) background from the target in the beampipe downstream of the target. The 2 m long region between the first collimator and the second (acceptance-defining) collimator, each 15 cm thick, was completely surrounded by thick concrete shielding. A resistive toroidal spectrometer magnet was situated just downstream of the third 11-cm-thick cleanup collimator.

The spectrometer magnet consisted of eight coils supported by an aluminum support structure in the shadow of the collimator, not intruding into the scattered electron acceptance.   Magnetic fields between the coils generated a toroidal field around the beam axis along the 2.2 m length of the magnet coils, which bent scattered electrons radially outward. The $\int{B\, dl}$ was about 0.9 T-m at the average electron scattering angle of $7.7^\circ$. The magnet was designed to separate elastic and inelastic events from hydrogen at the nominal detector location. However, such a design ($\Delta p/p \approx 10\%$) was inadequate to separate elastic and inelastic events from $A>$1 targets. As a result, corrections had to be made for inelastic backgrounds in this experiment.

A shielding hut was built around the experiment's detectors on all sides, including around the beam pipe which was shielded in lead. The borated concrete upstream wall of this hut had eight sculpted openings matching the shape of the scattered electron envelopes defined by the three upstream collimators and the magnet. The detectors consisted of 2-m-long rectangular bars of quartz 18 cm wide and 1.25 cm thick in the beam direction, arrayed symmetrically around the beam axis at a radius of $\approx 3.35$ m. Cherenkov light produced by scattered electrons   which reached  the detectors was read out at each end by 13-cm-diameter photomultiplier tubes (PMT) equipped with low-gain bases. Lead pre-radiators, 2-cm-thick, provided amplification of the scattered electron signal as well as suppression of soft backgrounds. 

Retractable and rotatable tracking drift chambers and trigger scintillation counters were employed just upstream of the detectors during dedicated periods with low beam current ($\approx 100$ pA - 1 nA) to measure the average four-momentum transfer $Q^2$, to benchmark simulations of the apparatus,  and to establish light-weighted acceptance corrections.

The apparatus described above was ideally suited for precise measurements of both the longitudinally and transversely polarized  parity-violating asymmetries on $^1$H ~\cite{Androic:2018kni, Androic:2020rkw}. However, it was not ideal for the study of $A>1$ nuclei reported here. Previous experiments on  $A>1$ nuclei used high-resolution magnetic spectrometers to isolate the elastically scattered electrons from the nuclear excited states, other target alloy elements, quasi-elastic scattering, and the inelastic $e N \rightarrow e^\prime \Delta$ reaction. In this experiment the contributions from these processes could not be isolated from the measured asymmetries, and instead had to be estimated and corrected for. In the following sections, these corrections and how they were estimated will be discussed in detail.

The data were obtained in four distinct data sets. 
In the first,  the beam was transversely polarized in the horizontal orientation, the beam current was 75~$\mu$A and a total of 1.6 C of integrated beam current was incident on the $^{12}$C target. The
three remaining data sets were obtained using the $^{27}$Al target. In the first of these, the beam polarization was in the vertical direction, the beam current was 24~$\mu$A, and the integrated beam current was 0.5~C. In the remaining two data sets, the beam current was increased to 61~$\mu$A, and a total of 3.3~C of integrated beam current was delivered, split approximately equally between a data set with horizontal orientation and a set with vertical orientation of the beam polarization.

A comprehensive GEANT4~\cite{Agostinelli:2002hh} simulation of the experimental apparatus was developed, benchmarked with measurements using the tracking system~\cite{Pan:2016rbx}, and was used for acceptance and radiative corrections as well as subtraction of various physics backgrounds, as discussed below. 
\section{\label{sec:analysis}Data Analysis}

In this section the corrections made and procedures used to determine the beam-normal single-spin asymmetry $B_n$ for each target are described. These include  corrections to the asymmetry data for spin-correlated fluctuations in the beam properties, fitting the angular dependence of these data in order to extract the amplitude of the azimuthal variation, and corrections for various backgrounds and other effects. 
Further details of this data analysis
can be found in Ref.~\cite{marty-thesis} for the $^{12}$C data and  Ref.~\cite{kurtis-thesis} for the  $^{27}$Al data. The hydrogen BNSSA datum obtained from this experiment is described in Ref.~\cite{BWaidyawansa_phd,Androic:2020rkw}.

\subsection{Determination of individual detector asymmetries $A^i_{\rm msr}$} \label{sec:regression}

The signals from each end of the eight Cherenkov detectors were integrated for each $\uparrow$ and $\downarrow$ spin state of the beam. The resulting averaged detector $i$ ($i = 1,8$) asymmetries were calculated for each 
quartet spin-pattern using 
\begin{equation}
A^i_{\rm raw} = \frac{Y^i_\uparrow - \,Y^i_\downarrow}{Y^i_\uparrow + \,Y^i_\downarrow} 
\label{eq:raw}
\end{equation}
where $Y^i_{\uparrow (\downarrow)}$
is the charge-normalized detector yield for detector $i$ in the $\uparrow \! \! (\downarrow)$ spin state, after subtraction of the electronic pedestal. $Y^i_{\uparrow (\downarrow)}$ was 
summed over the two windows of the same spin-state in each quartet.

 For each detector $i$, and for each quartet, false asymmetries in $A^i_{\rm raw}$ due to spin-correlated variations in the beam properties were corrected for using 
 \begin{equation}
 \label{eq:regression}
A^i_{\rm msr} = A^i_{\rm raw} - \sum\limits^5_{j=1} \left( \frac{\partial A^i}{\partial \chi_j} \right) \Delta \chi_j 
  \end{equation}
 where $\Delta \chi_j$ are the measured spin-correlated differences in beam trajectory or energy over each spin quartet, and the sensitivities $\partial A^i / \partial \chi_j$ were  determined using multi-variable linear regression. 
The natural random fluctuations in the trajectory and energy of the beam during the course of the measurement were large enough to enable these sensitivities to be extracted with sufficient precision for these corrections.

 The measured asymmetry in each detector
 was then corrected for two additional 
 effects: (i) the averaging of the azimuthally-varying asymmetry over the light-weighted angular acceptance of an individual detector, and (ii) any non-linear response of the detector to changes in yield. 
 The factor $R_{\rm av}$ 
accounts for averaging of the asymmetry over the effective azimuthal acceptance ($\approx$ 22$^{\circ}$) of a given Cherenkov detector~\cite{Androic:2013rhu}. In the ideal case of 100\% beam polarization, an individual detector centered at an azimuthal angle of $\phi_0 $ with an 
angular acceptance covering $\pm \delta\phi$ would measure an asymmetry given by
\begin{equation}A_{\rm msr}(\phi_0) = \frac{B_n}{2\delta \phi}\int_{\phi_0-\delta\phi}^{\phi_0+\delta\phi}\sin(\phi)d\phi = B_n\langle \sin\phi_0 \rangle .
\label{eq:rav}
\end{equation}
Thus the measured asymmetries $A_{\rm msr}$ have to be scaled by the factor $R_{\rm av} = \frac{\langle \sin\phi_0 \rangle}{\sin\phi_0}$.
However the optical response of the sum of both ends of a given Cherenkov detector varies by typically 10\% along the length of each detector~\cite{ALLISON2015105}, and is thus a function of $\phi$. Therefore the integral in Eq.~\ref{eq:rav} was performed with the integrand weighted by each detector's measured optical response function, yielding $R_{\rm av} = 0.9862 \pm 0.0036$. An additional correction factor $R_l$ is used to account for the non-linearity in the Cherenkov detector readout chain (photomultiplier tube, low-noise voltage-to-current preamplifier, and analog-to-digital converter) as described in Ref.~\cite{ALLISON2015105}. Bench studies using light-emitting diodes were conducted in order to determine any non-linearity in the response. At the signal levels appropriate to these two targets,
the non-linearity was found to be $0.14 \pm 0.50$\%~\cite{Duvall_phd}, so the correction factor was $R_l = 1.0014 \pm 0.0050$.

\subsection{Extraction of azimuthal asymmetry variation $A_{\rm exp}$}

For each of the targets, and for each of the four
data sets,  the measured asymmetries $A^i_{\rm msr}$ in each detector $i$ were sign-corrected for the presence
or absence of the IHWP at the electron source, averaged over the data set, and then fit to
\begin{equation}
\label{eq:bnsa_phi_dependance}
A^i_{\rm msr}(\phi_i) = 
 R_lR_{\rm av}B_{\rm exp}\sin(\phi_{\rm s}-\phi_i + \phi_{\rm off}) + C,
\end{equation} 
in order to extract the experimental asymmetry $B_{\rm exp}$.
Here  $\phi_{s}$ is the azimuthal angle of the electron polarization $\vec{P}$,  $\phi_i$ is the azimuthal angle of the $i^{th}$ detector in the plane normal to the beam axis, and $R_l$ and $R_{\rm av}$ were discussed above. The detector number $i$ corresponds to the azimuthal location of the detectors, starting from beam left (Detector 1) where $\phi_i=0^\circ$, and increasing clockwise every $45^\circ$. The values of
$B_{\rm exp}$ extracted from the fits are
presented in Table~\ref{tab:reg}. 

\begin{table*}[!hhtb]
    \centering
        \caption{Fitted asymmetries $B_{\rm exp}$ before (``raw") and after (``regressed") the linear regression correction to remove spin-correlated false asymmetries. The last two columns denote the phase $\phi_{\rm off}$ and offset $C$ from  Eq.~\ref{eq:bnsa_phi_dependance} obtained from fits to the regressed data. The data sets are labeled by the target and the direction of the beam polarization (horizontal or vertical). The uncertainties are statistical only. Also indicated are the $\chi^2$ per degree of freedom of each fit; there were five degrees of freedom in each fit. }
    \begin{tabular}{lccccccc}
  Data Set   &  $B_{\rm exp}$ (raw) & ~~$\chi^2/{\rm dof}$~   & $B_{\rm exp}$ (regressed) & ~$\chi^2/{\rm dof}$~ & regressed-raw & $\phi_{\rm off}$ (regressed)& $C$ (regressed)\\
 & (ppm) & & (ppm)&   & (ppm) & (radians) & (ppm) \\
  \hline 
  \rule{-3pt}{2.5ex}
 $^{12}$C Horizontal & $-8.57 \pm 0.61$ & 0.80 & 
 $-8.50 \pm 0.61$ & 0.81 & 0.07 & -0.099 $\pm$ 0.070 & 0.03 $\pm$ 0.43\\
\hline  \rule{-3pt}{2.5ex}
 $^{27}$Al Vertical \#1~~  & $-9.32 \pm 0.61$ & 1.16 & $-9.91 \pm 0.61$ &  1.16 & -0.59 & 0.114 $\pm$ 0.062 & 0.55 $\pm$ 0.43\\
 $^{27}$Al Horizontal&  $-8.54 \pm 0.51$ & 1.14 & $-8.60 \pm 0.50$ & 1.17 & -0.06& 0.053 $\pm$ 0.059 & -0.11 $\pm$ 0.35\\
 $^{27}$Al Vertical \#2 & $-8.02 \pm 0.74$ & 0.62& $-8.73 \pm 0.73$ & 0.64 & -0.71& -0.009 $\pm$ 0.084 & 0.20 $\pm$ 0.51\\
\hline  \rule{-3pt}{2.5ex}
  $^{27}$Al average & $-8.69 \pm 0.34$ & & $-9.04 \pm 0.34$  & & -0.35\\
         \hline
    \end{tabular}
    \label{tab:reg}
\end{table*}

A floating offset in phase $\phi_{\rm off}$ was included in the fit function (Eq.~\ref{eq:bnsa_phi_dependance}) to allow for possible position offsets of the detector in the azimuthal plane, and a floating constant $C$ was included to represent any background or false asymmetries which have no azimuthal variation. Such an azimuthally-symmetric asymmetry could arise due to, for example, the weak-interaction-induced parity-violating asymmetry, which could be generated by any residual longitudinal component to the beam polarization. For each of the data sets the fitted values for $\phi_{\rm off}$ and $C$ were consistent with zero, and the value of $B_{\rm exp}$ extracted was insensitive to the presence or absence of these two extra fit parameters. These findings for $\phi_{\rm off}$ and $C$ are also consistent with those from the precision result on hydrogen using the same apparatus, published earlier ~\cite{BWaidyawansa_phd,Androic:2020rkw}. For each target, the maximum deviation in $B_{\rm exp}$ between fits done with or without different combinations of 
$\phi_{\rm off}$ and $C$ was chosen as a ``fit function'' systematic uncertainty $B_{\rm fit}$, which was
$\pm 0.042$ ppm for $^{12}$C and $\pm 0.050$ ppm for $^{27}$Al.

\begin{figure}[!htbp]
\includegraphics[width=0.75\columnwidth]{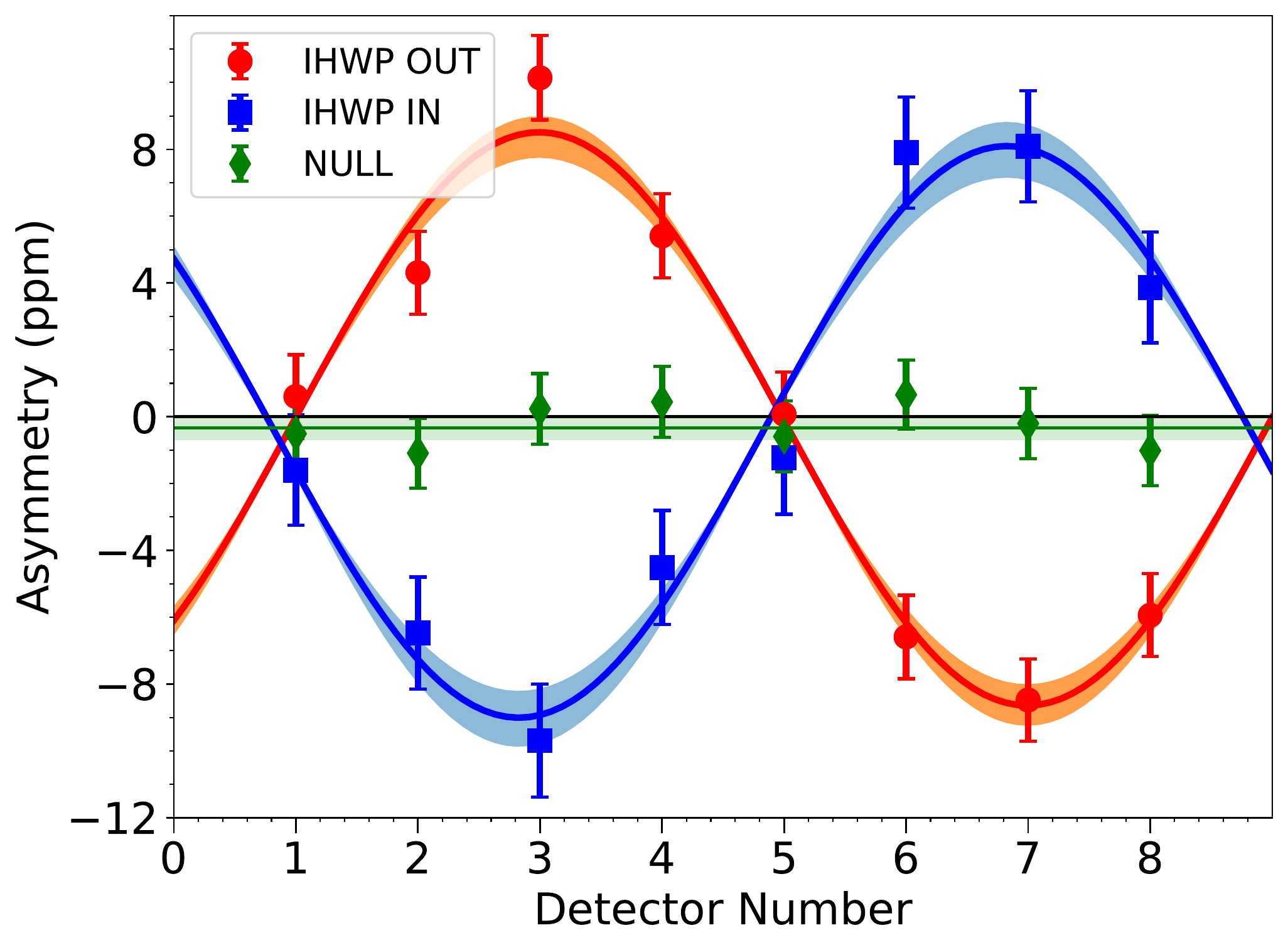}
\caption{Asymmetries measured with the IHWP IN (blue squares) and OUT (red circles) during the $^{27}$Al dataset with horizontal orientation of the transversely polarized beam, plotted vs. detector number. The detector number corresponds to the azimuthal location of the detectors, starting from beam left (Detector 1) where $\phi_i=0^\circ$, and increasing clockwise every $45^\circ$. These data include the regression correction discussed in Eq.~\ref{eq:regression}. The error bars shown are statistical only. The null asymmetries 
((OUT $+$ IN)/2) are denoted by the green diamonds. Solid lines correspond to fits using Eq.~\ref{eq:bnsa_phi_dependance} for each half-wave state, and to the average null asymmetry offset. The shaded areas represent the uncertainty in the fitted amplitudes for each IHWP orientation, as well as for the average null asymmetry  which is consistent with zero, as expected.
\label{fig:ihwp_nulls}
}
\end{figure}

A useful ``null" test for the presence of a certain class of false asymmetries is the behavior of the asymmetry under the ``slow spin reversal" accomplished using the  insertable half-wave plate.
In the absence of false asymmetry, the measured asymmetry should be equal in magnitude and opposite in sign for data taken with the IHWP inserted 
compared to that with the IHWP removed. 
Separate fits to the data with the IHWP inserted and data with the IWHP removed were done for each of the four data sets. In each case, the fitted $A_{\rm exp}$ was statistically consistent with the expected behavior. An example from one of the four datasets (horizontal $^{27}$Al) is shown in Fig.~\ref{fig:ihwp_nulls}.

To quantify the effect of the removal of spin-correlated false asymmetries on the extracted value of $B_{\rm exp}$, a similar
fit to that of  Eq.~\ref{eq:bnsa_phi_dependance} was also performed, but instead using the raw asymmetries 
$A^i_{\rm raw}(\phi_i)$. Table ~\ref{tab:reg} provides the results of the fits to both the
raw ($A^i_{\rm raw}$) and the linear-regression corrected asymmetries ($A^i_{\rm msr}$) for each data set. The linear regression corrections were comparable with the statistical uncertainty for the two vertical data sets, but were an order of magnitude smaller for the horizontal data sets.

Different choices could be made for the set of BPMs used to determine the beam trajectories in the linear regression corrections (Eq.~\ref{eq:regression}). Several different 
combinations of BPM selections were studied, and the largest deviation in the extracted values of $B_{\rm exp}$ was taken as a systematic uncertainty $B_{\rm reg}$ for each target. For $^{12}$C, $B_{\rm reg}$ was $\pm 0.002$~ppm and for $^{27}$Al it was  $\pm 0.020~$ppm. 

\begin{figure}[!htbp]
\includegraphics[width=\columnwidth]{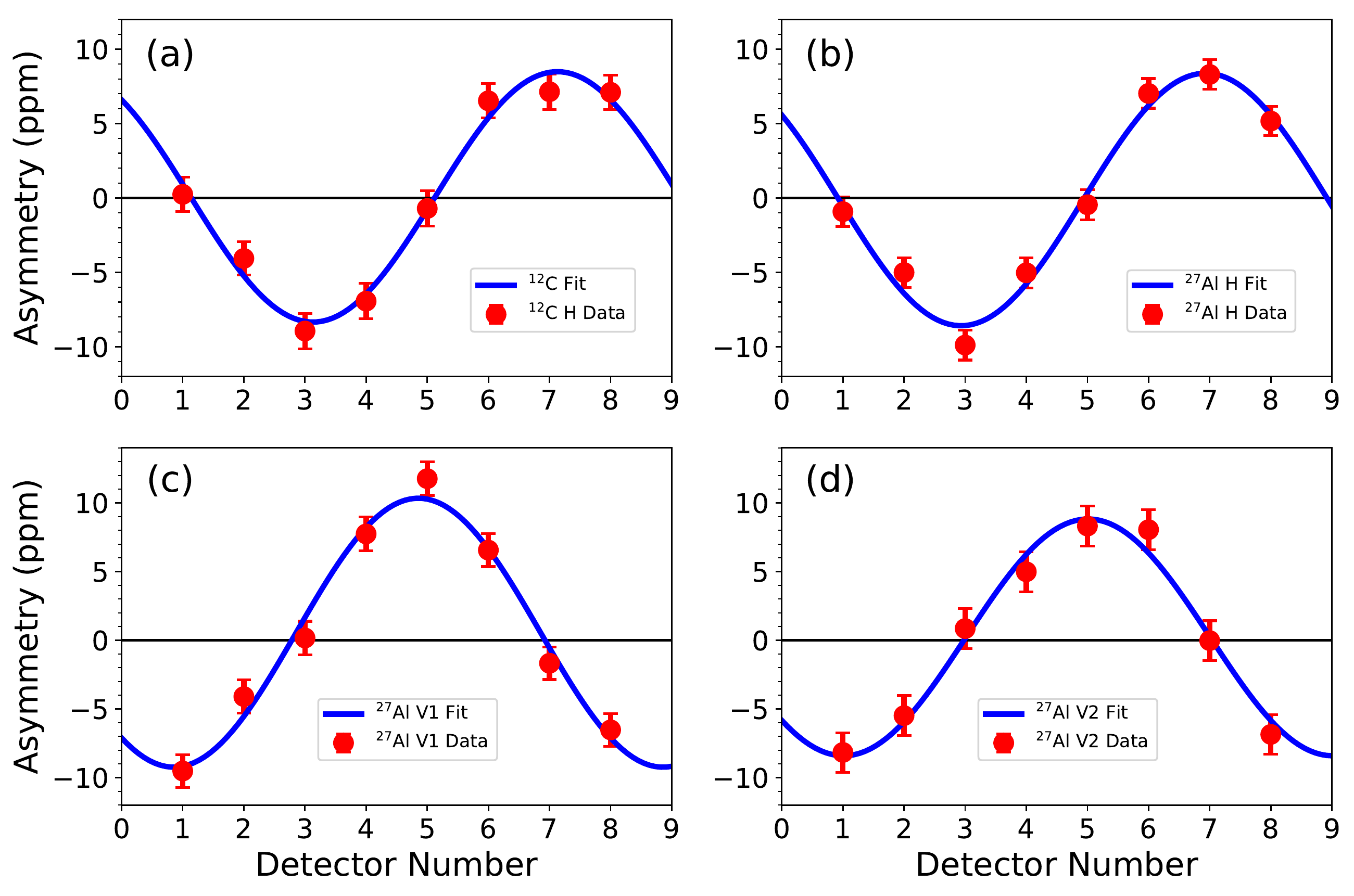}
\caption{The azimuthal asymmetry distributions measured in this experiment (red circles) and fits (blue lines) using Eq.~\ref{eq:bnsa_phi_dependance} are shown for the four datasets: $^{12}$C with horizontal transverse polarization (a), $^{27}$Al horizontal (b), $^{27}$Al vertical in run 1 (c), and $^{27}$Al vertical in run 2 (d). The abscissa denotes the detector number, as described in Fig.~\ref{fig:ihwp_nulls}. Uncertainties shown are statistical only.\label{fig:All4wiggles}
}
\end{figure}

The (regressed) data and the fitted azimuthal dependencies from which $B_{\rm exp}$ was determined for each of the four datasets are shown in Fig.~\ref{fig:All4wiggles}. Uncertainties shown are statistical only. Each fit consists of eight measurements and three free parameters, giving five degrees of freedom in each fit.

\vspace*{2ex}
 
 \subsection{Corrections to $B_{\rm exp}$:  acceptance, beam polarization, instrumental false asymmetry} \label{sec:Aexp_corrections}
 
In order to extract the beam-normal single-spin asymmetry $B_n$ from the measured, regressed $B_{\rm exp}$, corrections were made for beam polarization, radiative effects, several backgrounds, and an instrumental false asymmetry.  These corrections were applied using 
 \begin{equation}
\label{eq:asymmetry_correction_formula}
B_n = R_{\rm tot}\left[\frac{\frac{B_{\rm exp}}{P}- \sum_{i} f_iB_i}{1-\sum_{i} f_i}\right] + B_{\rm bias} \;\;  .
\end{equation}

Here $B_i$ is the background asymmetry generated by the $i^{th}$ background 
(quasielastic scattering, inelastic scattering, nuclear excited states,  neutral backgrounds, and alloy elements in the case of $^{27}$Al.) with fractional contribution to the detector signal $f_i$. The background
corrections will be discussed in the next section (\ref{Sec:backgrounds}). In the remainder of this section we discuss 
the  other components in Eq.~\ref{eq:asymmetry_correction_formula}.

\paragraph*{Beam Polarization:} The beam polarization $P$
was measured during longitudinal polarization data-taking conducted just before and after each transverse data set, using the M\o ller  and Compton polarimeters~\cite{Hauger:1999iv,Magee:2016xqx,Narayan:2015aua} in Hall C. For the $^{12}$C data set the beam polarization was $P = 0.8852 \pm 0.0068$. For the three $^{27}$Al data sets the (asymmetry and statistics-weighted) average value was $P = 0.8873 \pm 0.0071$. 
 During the transverse running, the degree of transverse polarization was intermittently measured via several null measurements with the M\o ller polarimeter, which is only sensitive to longitudinal beam polarization.  The worst case found (during a horizontal transverse running period) was  a  $2.19 \pm  0.26 \%$ residual longitudinal polarization, indicating the degree of transverse polarization was $\geq \sqrt{1-(2.19/88.73)^2} = 99.97 \%$ 
transverse. Other null checks made during the transverse running were consistent with zero residual longitudinal polarization.

\paragraph*{Radiative and acceptance corrections: }

The factor $R_{\rm tot} = 1.0054 \pm 0.0046$ is the product of several individually small (percent-level) corrections \cite{BWaidyawansa_phd}. These include electron energy-loss and depolarization from electromagnetic radiation (internal and external bremsstrahlung), and the non-uniform $Q^2$ distribution across the detectors coupled to variation in the  light-collection across the detectors. It also corrects for the fact that, due to the large detector acceptance, we measure $B_n(Q)$ over a range of $Q$. Since $B_n$ varies roughly linearly with $Q$, we need to correct the acceptance-averaged value $\langle B_n(Q)\rangle$ to the value that would arise from point scattering at the central $\langle Q \rangle$, {\em i.e.} $B_n(\langle Q \rangle)$. The uncertainty in $R_{\rm tot}$ also accounts for the uncertainty in the central value of the acceptance-averaged $\langle Q \rangle$. The procedures used to determine these individual corrections follow those described in~\cite{Androic:2018kni} for our $A_{\rm PV}$ measurement on the proton (where $R_{\rm tot} = 0.976 \pm  0.008$), but with slightly different numerical results due to the use of different targets, and the fact that $A_{\rm PV}$ varies linearly with  $Q^2$ instead of with $Q$ in the case of $B_n$. 

\paragraph*{Rescattering Bias}

An instrumental false asymmetry, the rescattering bias, $B_{\rm bias}$, was accounted for as described in detail in \cite{Androic:2018kni}.  The transversely-polarized electrons, scattered from the target, retained much of their transverse polarization as they were transported through the  spectrometer magnet. Lead pre-radiators were located in front of each 
of the Cherenkov detectors~\cite{ALLISON2015105} 
to amplify the electron signal and help suppress soft backgrounds. When these polarized electrons showered in the pre-radiators, they could be reduced in energy enough that the analyzing power due to low-energy Mott scattering was sufficiently large to cause measurable asymmetries. The false asymmetry in the present case of transversely polarized beam was larger than it was for longitudinal polarization, {\em i.e.} for the weak charge measurement
~\cite{Androic:2018kni}. This is because, for longitudinal polarization, the analyzing power in the pre-radiator leads to an asymmetry  of equal magnitude and opposite sign for the signals detected in the two PMTs on either end
of each Cherenkov detector. This largely canceled when the signals were summed in the data analysis. In the case of transversely polarized beam, the analyzing power affects both PMTs identically, so there is not a similar cancellation. Instead this generated an azimuthally-varying false asymmetry $B_{\rm bias} = 0.125 \pm 0.041$~ppm. 
$B_{bias}$ is a false asymmetry across each detector bar that would be present for $B_n$ even in a perfectly symmetric identical array of detectors with no imperfections.
\subsection{Corrections to $A_{\rm exp}$: backgrounds}\label{Sec:backgrounds}
 
\paragraph*{Alloy elements:} 

The $^{27}$Al target was not made from pure aluminum; rather, it was an alloy (7075-T651) containing  89.2\% Al by weight, 5.9\% Zn, 2.6\% Mg, 1.8\% Cu, and $\approx$0.6\% other elements. The reason this alloy was chosen instead of pure aluminum was its superior strength; the ultimate tensile strength is 572 MPa versus 90 MPa for pure aluminum. It could thus be used to make much thinner windows for the liquid-hydrogen target cell deployed in the weak charge measurement \cite{Androic:2018kni}, with correspondingly less background. 
Because the elemental composition of a given alloy can vary, asymmetries were measured from solid aluminum alloy targets composed of the same lot of material used for the hydrogen target cell windows in order to characterize the background from the target cell in the weak charge measurement. In order to also report a result for $^{27}$Al in the work described here, however, it is necessary to subtract the small contributions from the alloy elements other than aluminum. The fractional contributions $f_i$ to the detected yield from each alloy element were determined through simulation. In the simulation, only the elastic scattering cross section (which dominates at the small-angle kinematics of this experiment) was considered for the alloy elements. The elastic cross sections for $^{27}$Al and the six most abundant elements in the alloy ($^{64}$Zn, $^{24}$Mg, $^{63}$Cu, $^{52}$Cr, $^{56}$Fe, and $^{28}$Si)
were calculated  by Horowitz and Lin~\cite{chuck} using a relativistic mean-field model and including the effects of Coulomb distortions; a 10\% uncertainty was estimated for each cross section. For the remaining two alloy elements (Mn and Ti) the cross sections were estimated using form factors extracted from experimental Fourier-Bessel coefficients~\cite{DEVRIES1987495}, and a 50\% uncertainty was assumed for both cross sections. The estimated fractional contributions are given in Table~\ref{tab:alloy}; the total fraction of the experimental yield arising from the alloy elements was $f_{\rm alloy} = 5.41 \pm 0.34$\%. 

There have yet to be $B_n$ measurements made for any of these alloy elements, with the exception of $^{28}$Si. In order to estimate the beam-normal single-spin asymmetry $B_i$ for each alloy element, we assumed the scaling of Eq.~\ref{eq:scaling}, with $Q^2 = 0.0237$ GeV$^2$, and $\widehat{B}_n  = -33.0 \;{\rm ppm}/{\rm GeV}$, where the value of $\widehat{B}_n$ was taken from our published result on the proton \cite{Androic:2020rkw} at essentially the same kinematics as the present measurements. These estimated $B_i$ are tabulated in Table~\ref{tab:alloy}. In the case of $^{28}$Si, the A1 collaboration at Mainz has reported a result for $B_n$ ~\cite{Esser:2020vjb}, albeit at  different kinematics than for the present measurement. Their $B_n$ agreed with the predictions for this nucleus using the model of Gorchtein and Horowitz to within about 30\%. Thus we ascribe a 30\% uncertainty to $B_n$ for each of the alloy elements. An average asymmetry for the alloy elements, weighted by the relative background fractions, was calculated to be $B_{\rm alloy}=- 10.7 \pm 2.0$ ppm.

The $^{12}$C target was elementally pure, so no corrections for alloy elements were required in that case. 

\begin{table}[!htb]
    \caption{Background fractions $f_i$ and beam-normal single-spin asymmetry $B_i$ estimates for the alloy elements
present in the aluminum target. The net background fraction and the weighted average background asymmetry from alloy materials are also listed.}
    \label{tab:alloy}    
    \centering
    \begin{tabular}{ccc}
\hline
Element & Background fraction ($f_i$) & Asymmetry ($B_i$)  \\
& (\%) & (ppm) \\
\hline
Zn & $2.375 \pm 0.249$ & $-11.0 \pm 3.3$ \\
Mg & $2.088 \pm 0.219$ & $-10.3 \pm 3.1$ \\
Cu & $0.683 \pm 0.073$ & $-11.1 \pm 3.3$ \\
Cr & $0.100 \pm 0.011$ & $-11.0 \pm 3.3$ \\
Si & $0.080 \pm 0.009$ & $-10.2 \pm 3.0$ \\
Fe & $0.054 \pm 0.006$ & $-10.9 \pm 3.3$ \\
Mn & $0.018 \pm 0.009$ & $-11.1 \pm 3.3$ \\
Ti & $0.014 \pm 0.007$ & $-11.0 \pm 3.3$ \\
\hline
net Alloy & $5.41 \pm 0.34$ &  $-10.7 \pm 2.0$ \\
 \end{tabular}
\end{table}

\paragraph*{Neutral backgrounds:}
A small fraction of the detector yield was due to neutral events (predominantly soft gammas). These neutral particles arose due to both the primary electron beam interacting in various beamline elements, including a tungsten beam collimator~\cite{ALLISON2015105}, and to the scattered electrons interacting in the triple collimator system or in the spectrometer magnet 
structure. These backgrounds were carefully 
studied for the weak charge measurement, as detailed in Ref.~\cite{Androic:2018kni}. Similar studies were done for the $^{27}$Al target~\cite{kurtis-thesis}, which found a total
neutral contribution to the yield of $f_{\rm neut} = 0.69 \pm 0.45$\%. The same contribution was applied to the $^{12}$C target. In the absence of any measurement of an azimuthal asymmetry associated with these neutral events, we conservatively assume  $B_{\rm neut} = 0 \pm 10$ ppm for both targets.

\paragraph*{Pions:}
A $\pi^-$ background from the proton target in the $Q_{\rm weak}$ $A_{PV}$ measurement was only possible if two or more pions were produced. Those pions mostly fell outside even the wide momentum acceptance of the $Q_{\rm weak}$ apparatus. However, single $\pi^-$ production from the neutrons in $^{27}$Al was possible. Simulations were performed~\cite{kurtis-thesis} using the Wiser~\cite{Wiser} pion production code on protons and neutrons scaled to $^{27}$Al but neglecting nuclear medium effects. The result was $f_{pion}=0.06\%$, so small that no correction was necessary for either $^{27}$Al or $^{12}$C.

\subsubsection{Corrections to $A_{\rm exp}$: Background from non-elastic physics processes}\label{Sec:non-elastic}

The Q$_{\rm weak}$ spectrometer had a rather large acceptance bite (of order 150 MeV) in scattered electron energy $E'$.  This meant that, along with the desired elastically-scattered electrons, events were accepted from various non-elastic scattering processes. These included low-lying nuclear excited states, the giant dipole resonance (GDR), quasi-elastic scattering from individual nucleons, and inelastic  scattering from individual nucleons (pion production). All these processes, unresolved from the elastic-scattering peak, contributed to the measured detector yield $Y_{\uparrow (\downarrow)}$ used in the asymmetry analysis. For each of these processes, the fractional contribution to the measured yield was estimated using simulation, and the asymmetry associated with each process was estimated from previous measurements or theoretical expectations, as described below. 
 
\paragraph*{Quasielastic and inelastic scattering:}

The fractional contribution to the yield arising from quasielastic scattering $f_{\rm QE}$ and inelastic scattering $f_{\rm inel}$ was determined for each target by simulation. 
The quasielastic and inelastic cross sections used in the simulation were obtained from an empirical fit~\cite{Christy_future} to world data on inclusive electron-nucleus scattering, including both $^{12}$C and $^{27}$Al.  The fit was based on the picture of scattering from independent nucleons in the impulse approximation.  The quasielastic contribution was modeled using input parameterizations of the nucleon form factors extracted from cross section data with the smearing due to Fermi motion of the nucleon in the nucleus accounted for by utilizing the super-scaling formalism of Donnelly and Sick~\cite{donnelly-sick:99}.  The inelastic contribution was accounted for by utilizing nucleon-level cross sections determined from fits to inclusive scattering from proton and deuteron targets with a Gaussian smearing to account for the Fermi motion, and medium modification factors to account for the EMC effect.   The fit utilized the approach of Bosted and Mamyan \cite{Bosted:2012qc}, but included a number of improvements in the kinematic region relevant for the current analysis at low $Q^2$ and $W$.   In particular, Bosted and Mamyan only included data with $Q^2 > 0.2$ and introduced an {\it ad hoc} medium modification factor to the nucleon magnetic form factor to help improve the comparison to the cross section data.  In the region unconstrained by data at very low $Q^2$ this resulted in a significant suppression of the quasielastic cross section, with this strength then absorbed into an empirical contribution associated with 2-body contributions such as meson-exchange currents (MEC).  In contrast, the new fit included data down to $Q^2 = 0.045$ and was able to improve the description of the data across the kinematic region of the fit without the need for modification of the nucleon form factors. The agreement between the fit and the total cross section data, in the low-$Q^2$ region relevant to the present experiment, was typically at the 5-10\% level. An additional 10\% uncertainty was added in quadrature to represent the ability of the fit to separate the quasielastic from the inelastic processes in this region.
The estimates of contributions from 2-body effects in the new version of the model were much smaller than either the quasielastic or inelastic processes, and were therefore neglected. The extracted fractional yield estimates (inside the acceptance of the experiment) were $f_{\rm QE} = 21.2 \pm 2.9 \%$ and $f_{\rm inel} =0.66 \pm 0.10 \% $  ($^{27}$Al) and  $f_{\rm QE} = 15.9 \pm 2.2 \%$ and $f_{\rm inel} = 0.40 \pm 0.06 \%$ ($^{12}$C).

Electromagnetic quasi-elastic interactions at the small angles and small momentum transfers of the current experiment are dominated by scattering from the proton. Therefore, the beam-normal single-spin asymmetry for quasielastic scattering $B_{\rm QE}$ was estimated using the 
$B_n$ result for elastic scattering from the proton  measured by our collaboration at the same kinematics using the same apparatus ~\cite{BWaidyawansa_phd,Androic:2020rkw}: $B_n=-5.194 \pm 0.106$~ppm. The uncertainty on this asymmetry was increased to $\pm 1$~ppm to account for the possibility of nuclear medium effects and for the neglect of quasielastic scattering from the neutron.

The beam-normal single-spin asymmetry for the hadronic  inelastic events $B_{\rm inel}$ was estimated using our data from a separate measurement~\cite{Nuruzzaman_phd,Nuruzzaman:2015vba}. In that measurement, with the electron beam polarized in a transverse direction, the spectrometer magnetic field was reduced to 75\% of its nominal strength, thereby bringing electrons scattered in the $ep \rightarrow e'\Delta^+ $ process onto the Cherenkov detectors. Analyzing those data using a similar method to that described here led to the preliminary result $B_{\rm inel} = 43.0 \pm  16.0$  ppm ~\cite{Nuruzzaman_phd,Nuruzzaman:2015vba}. Note that the observed sign of the asymmetry is opposite to that of the asymmetry for elastic scattering; this sign difference was predicted theoretically ~\cite{Carlson:2017lys}.
We have assumed that this same asymmetry also applies to the inelastic process on the neutron, and that there are no significant nuclear medium modifications, and so the same $B_{\rm inel}$ was assigned for both the $^{12}$C and the $^{27}$Al targets. 

\paragraph*{Nuclear Excitations:} 
Electroexcitation of the low-lying discrete excited states of $^{27}$Al has been studied in several experiments~\cite{Singhal:1977wvn,Hicks:1980bv,Ryan:1983zz}, and form factors extracted. The eleven states with the largest form factors in the range of the present experimental acceptance 0.68 fm$^{-1}$ $< Q < 1.20 \; {\rm fm}^{-1}$ were considered here. Differential cross sections were calculated using these form factors and the fractional yield from each of these states determined from simulation~\cite{kurtis-thesis}. The results are
presented in Table~\ref{tab:al_excited}. 

Experimental electroexcitation form factors for the low-lying excited states of $^{12}$C are also available.  Differential cross sections were calculated for the three states with the largest form factors in the experimental acceptance~\cite{Crannell:1964zz,
PhysRev.148.1107,PhysRevLett.27.745}, and the 
fractional yields from each of these states determined from simulation~\cite{marty-thesis}. The results are presented in Table~\ref{tab:c_excited}. 

\begin{table}[]
    \centering
        \caption{Simulated fractional contributions from unresolved nuclear excitations for $^{27}$Al. The 2.990 MeV represents an  unresolved doublet of states. The asymmetries for these states were taken to be the $^{27}$Al elastic $B_n$ asymmetry $\pm 100\%$.}
    \begin{tabular}{ccc}
 Energy &  $J^{P}$ & Background fraction ($f_i$) \\
(MeV) & & (\%) \\
 \hline 
0.844
&  $1/2^+$ & $0.27 \pm 0.04$ \\
1.014 &  $3/2^+$  &  $0.41 \pm 0.10$ \\
2.211 &  $7/2^+$ & $1.35 \pm  0.16$ \\
2.735 &  $5/2^+$  &  $0.19 \pm 0.02$ \\
2.990 &  $3/2^+$ &  $0.93 \pm 0.07$ \\
4.540  &    &  $0.06 \pm 0.01$ \\
4.812 &  $5/2^+$  &  $0.09 \pm 0.02$ \\
5.430 &    &  $0.17 \pm 0.03$ \\
5.668 & $9/2^+$   &  $0.08 \pm 0.02$ \\
7.228&  $9/2^+$  &  $0.18 \pm 0.06$ \\
7.477&    &  $0.10 \pm 0.07$ \\
21 &    $1^-$ (GDR) &  $0.045 \pm 0.022$ \\
\hline
Total & & $ 3.88 \pm 0.23$ \% \\
    \end{tabular}
    \label{tab:al_excited}
\end{table}

\begin{table}[]
    \centering
        \caption{Simulated fractional contributions from unresolved excitations for $^{12}$C. All other states contribute $<1\%$. The asymmetries for these states were taken to be the $^{12}$C elastic $B_n$ asymmetry $\pm 100\%$.
    }
    \begin{tabular}{ccc}
 Energy &  $J^{P}$ & Background fraction ($f_i$) \\
(MeV) & & (\%) \\
 \hline 
4.44 &  $2^+$    & $2.86 \pm 0.29$ \\
7.65 &  $0^+$    &  $0.92 \pm 0.09$ \\
9.64 &  $3^-$    & $0.93 \pm 0.09$ \\
24  &  $1^-$ (GDR) &  $0.077 \pm 0.038$ \\
\hline
 Total & & $ 4.71 \pm 0.31$ \% \\   \end{tabular}
    \label{tab:c_excited}
\end{table}

For each target, the yield contribution from the GDR was simulated using the Goldhaber-Teller model ~\cite{GOLDEMBERG1963242} with energy and width parameters taken from photoabsorption data~\cite{Varmalov2012,Ahrens:1975rq}. The simulated yield fraction $f_{\rm GDR}$ 
for $^{27}$Al was $0.045 \pm 0.022$\% and for $^{12}$C it was $0.077 \pm 0.038$\%
(see  Table~\ref{tab:al_excited} and  Table~\ref{tab:c_excited}). In total, the background fraction from the GDR and the low-lying excited states were $f_{\rm nucl} = 3.88 \pm 0.23$\% of the yield ($^{27}$Al) and 
 $f_{\rm nucl} = 4.71 \pm 0.31$\% of the yield ($^{12}$C).

There are neither theoretical calculations nor experimental measurements of $B_n$ for the low-lying nuclear excitations or the GDR, for either $^{12}$C or $^{27}$Al. The reactions exciting these states have only modestly different kinematics than the elastic scattering reaction of interest. Therefore we assume the same $B_n$ for these excited state transitions as for the elastic reaction for the given nucleus, but assign a conservative 100\% uncertainty to these values.

\section{\label{sec:results}Results}
The beam energy for both targets was $1.158 \pm 0.001$ GeV. The kinematics for the experiment were determined from a GEANT4 simulation, as benchmarked using the tracking chambers~\cite{Pan:2016rbx}.
The central kinematics (averaged over the acceptance) for 
$^{12}$C was $\langle Q^2 \rangle = 0.02516 \pm 0.0001$ GeV$^2$ ($Q=0.159$) and $\langle \theta_{\rm Lab}\rangle = 7.86^{\circ} \pm 0.02^{\circ}$, and for 
$^{27}$Al was $\langle Q^2 \rangle = 0.02357 \pm  0.0001 $ GeV$^2$ ($Q=0.154$) and 
$\langle \theta_{\rm Lab}\rangle = 7.61^{\circ} \pm 0.02^{\circ}$.

Summing the various backgrounds contributing to the detector yield  discussed in Sec.~\ref{Sec:backgrounds} for each target, we have $\sum_{i} f_i = 22.8 \pm 2.5 \%$ for $^{12}$C and
$\sum_{i} f_i = 31.8 \pm 3.0 \%$ for $^{27}$Al.
After all corrections were applied using Eq.~\ref{eq:asymmetry_correction_formula}, the resulting beam-normal single-spin asymmetries were $B_n =  -10.68 \pm  0.90 {\rm (stat)} \pm 0.57 {\rm (syst)}$  ppm for $^{12}$C and
$B_n =  -12.16 \pm  0.58 {\rm (stat)} \pm 0.62 {\rm (syst)}$  ppm for $^{27}$Al. Each of the corrections applied are tabulated in Table~\ref{Tab:Corrections}, as is the fractional contribution each correction made to the uncertainty of the $B_n$ values.  The  dilutions in Table~\ref{Tab:Corrections} for the discrete  nuclear state backgrounds (including the GDR), as well as the $^{27}$Al alloy background,  represent the sum of the relevant individual constituent  dilutions. The asymmetry uncertainties for these two composite backgrounds were rolled up in the table by dividing the quadrature sum of the individual $f_i B_i$ uncertainty contributions by the sum of the relevant dilutions.

\begin{table*}[!htb]
 \caption{Corrections applied to the  measured asymmetry $B_{\rm exp}$ in order to determine $B_n$ (see Eq.~\ref{eq:asymmetry_correction_formula} and text), and their contributions to the systematic uncertainty on $B_n$.   }
 \label{Tab:Corrections}
    \centering
    \begin{tabular}{lcccc }
\hline
   Quantity & Value  & Value  & $\Delta B_n/B_n$ (\%) & $\Delta B_n/B_n$ (\%) \\
      &    $^{12}$C & $^{27}$Al & $^{12}$C & $^{27}$Al  \\
   \hline 
 & & & & \\
  $P$:   Beam Polarization  &   $0.8852 \pm 0.0068$ & $0.8872 \pm 0.0070$ & 0.9 & 1.0 \\
   $R_{\rm tot}$: Kinematics \& Radiative effects & $ 1.0054 \pm 0.0046$  & $1.0054 \pm 0.0046$  & 0.5 & 0.5\\
   $R_{\rm av}$: Acceptance averaging &  $0.9862 \pm 0.0036$
   &  $0.9862 \pm 0.0036$ & 0.4 & 0.4\\
 $R_l$: Electronic non-linearity & $1.0014 \pm 0.0050$ &  $1.0014 \pm 0.0050$ & 0.6 & 0.6 \\   
 $B_{\rm fit}$: Fitting  & $0 \pm 0.042$ ppm &  $0 \pm 0.050$ ppm & 0.6 & 0.6\\
 $B_{\rm reg}$: Linear Regression & $0 \pm 0.002$ ppm & $0 \pm 0.020$ ppm & $<0.1$ & 0.3\\
$B_{\rm bias}$: Rescattering Bias & $0.125 \pm 0.041$ ppm & ~~$0.125 \pm 0.041$ ppm & 0.6 & 0.6 \\
 $f_{\rm neutral}$: & $0.69\pm 0.45$ \% & $0.69\pm 0.45$ \%  & 0.8 & 0.7 \\
$B_{\rm neutral}$:& $0\pm 10$ ppm &$0\pm 10$ ppm & 0.6 & 0.8 \\
 $f_{\rm alloy}$: & --- & $5.41 \pm 0.34$ \% & --- &  $<0.1$ \\
 $B_{\rm alloy}$: & --- & $-10.7 \pm 2.0$ ppm & --- & 1.3\\
$f_{\rm QE}$: & $15.9 \pm 2.2$ \% & $21.2 \pm 2.9$ \%& 1.5 &  2.4 \\
$B_{\rm QE}$:& $-5.2 \pm 1.0$ ppm & $-5.2 \pm 1.0$ ppm & 2.0 & 2.6 \\
$f_{\rm inel}$: & $0.40\pm 0.06$ \% & $0.66 \pm 0.10$ \% & 0.4 & 0.7 \\
$B_{\rm inel}$:& $43 \pm 16$ ppm & $43 \pm 16$ ppm & 0.8 & 1.3\\
$f_{\rm nucl}$ & $4.71 \pm 0.31$ \% & $3.88 \pm 0.23$ \% & $<0.1$ & $<0.1$\\
$B_{\rm nucl}$ & $-10.5 \pm 10.5$ ppm & $-12 \pm 5.5$ ppm & 3.9  & 2.6\\
        \hline
Total Systematic & & & 5.3 \%  & 5.2 \% \\
    \end{tabular}
\end{table*}

\section{\label{sec:discussion}Discussion}

In this section four different prisms are used to interrogate the new results on $^{12}$C and $^{27}$Al. First the data are compared to theoretical predictions made at the kinematics of this experiment using the optical model approach. Next, a global set of the world's BNSSA data is assembled in Table~\ref{tab:Globaldata}, which excludes backward angle data as well as neutron data from quasi-elastic scattering on the deuteron. 
 This global dataset is then  scaled linearly to the $Q$ of this experiment and plotted against atomic mass number $A$.  The same BNSSA dataset is then scaled instead by $Z/A$ and plotted against $Q$ to look for trends which may provide insight into the global behaviour of the BNSSA. Finally, a plot against $Q$ is made of $\widehat{B}_n$ derived from the global dataset using Eq.~\ref{eq:scaling}, from which even more insights are obtained.

\subsection{\label{sec:Gorch}Comparison to calculations}
\begin{figure}[!htbp]
\includegraphics[width=0.75\columnwidth]{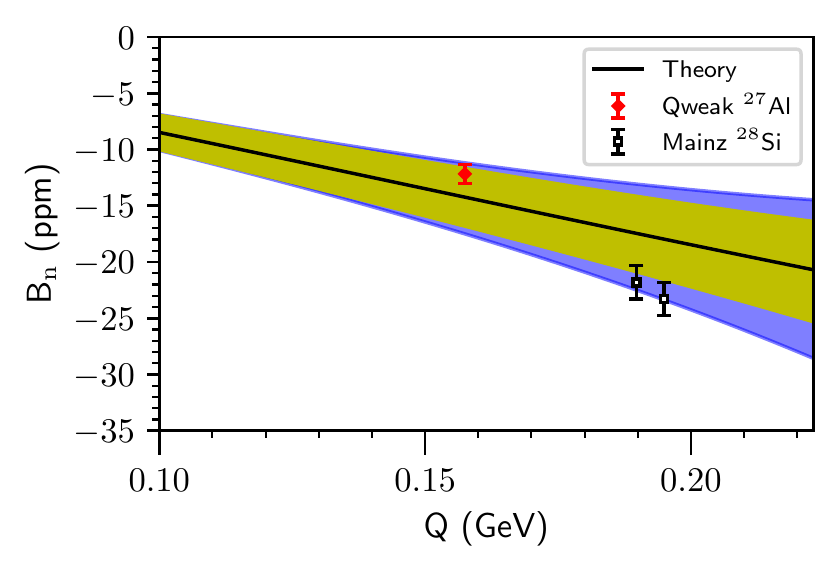}
\caption{\label{fig:aluminum_result}
The BNSSA measured in this experiment from $^{27}$Al (red diamond)  compared to predictions (black curve) by Gorchtein and Horowitz~\cite{Gorch} for  $Q=0.154$ GeV and $7.6^\circ$. The   error bars represent the quadrature sum of the statistical and systematic uncertainties. Recent data~\cite{Esser:2020vjb} from Mainz A1 on a neighboring nucleus, $^{28}$Si, are also included for comparison (open squares). 
The inner (yellow) and outer (blue) bands correspond to 10\% and 20\% uncertainties for the Compton slope parameter used in the calculation (see text). }
\end{figure}

\begin{figure}[!htbp]
\includegraphics[width=0.75\columnwidth]{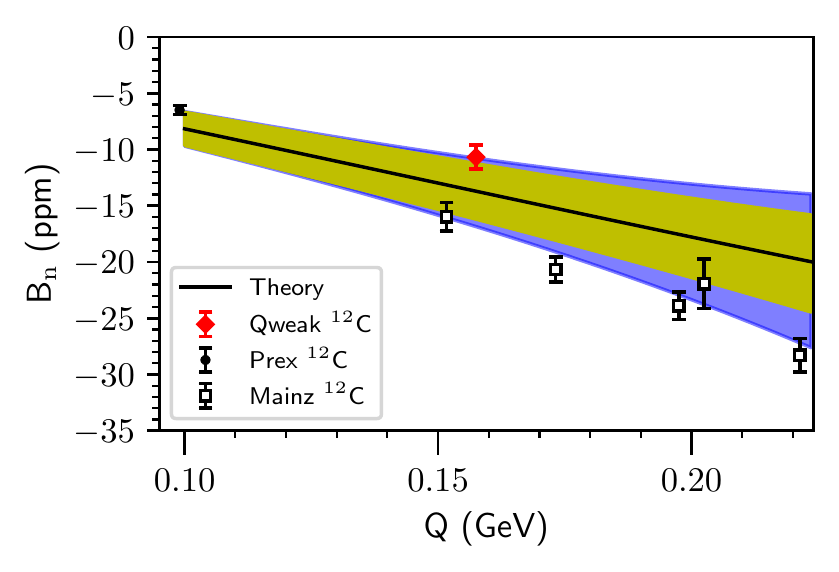}
\caption{\label{fig:carbon_result} The world $B_n$ data on $^{12}$C, including the result from this experiment (red diamond). The precise result from PREX \cite{Abrahamyan:2012cg} at the lowest $Q$ is also shown (circle), as well as five larger-angle results (open squares) from Mainz A1 \cite{Esser:2018vdp}.
The   error  bars  represent  the  quadrature  sum  of  the  statistical and systematic uncertainties. The inner (yellow) and outer (blue) bands correspond to 10\% and 20\% uncertainties for the Compton slope parameter used in the calculation (see text).}
\end{figure}

The present results are compared to predictions obtained in the optical model approach of  Gorchtein and Horowitz~\cite{Gorchtein:2008dy},
extended to the relevant nuclei~\cite{Gorch},
 in Fig.~\ref{fig:aluminum_result} (for $^{27}$Al) and Fig.~\ref{fig:carbon_result} (for $^{12}$C).
No other data exist for $^{27}$Al, however data on a neighboring nucleus ($^{28}$Si) have recently been published~\cite{Esser:2020vjb} at $E=0.570$ GeV near $20^\circ$. Although these calculations were made specifically at the kinematics of the Q$_{\rm weak}$ experiment~\cite{Gorch}, the $^{28}$Si results from Mainz are included in Fig.~\ref{fig:aluminum_result} for comparison.  Data from other  experiments on $^{12}$C at slightly different kinematics are also included in  Fig.~\ref{fig:carbon_result} along with the 
$^{12}$C result from this experiment. The PREX datum~\cite{Abrahamyan:2012cg} is another far-forward experiment ($\theta=5^\circ$, E=1.063 GeV, $Q^2=0.00984$ GeV$^2$). The Mainz data ~\cite{Esser:2018vdp} shown are at somewhat larger angles ($15.1^\circ \leq \theta \leq 25.9^\circ$, E=0.570 GeV, $0.023 \leq Q^2 \leq 0.049$ GeV$^2$).

The uncertainty bands shown for the calculations in both  Figs.~\ref{fig:aluminum_result} and \ref{fig:carbon_result} arise from two components, added in quadrature: the Compton slope parameter $B_c$, and terms not enhanced by the large logarithm $\ln(Q^2/m_e^2)$ (where $m_e$ is the electron mass)~\cite{Gorchtein:2008dy}. 
The inner and outer uncertainty bands in the figure arise from assigning either a 10\% or 20\% uncertainty to $B_c$, and the non-log-enhanced terms~\cite{Gorchtein:2008px} are assigned 100\% uncertainty. It should be noted that these calculations 
do not include the small elastic intermediate state contribution, 
only the (dominant) inelastic intermediate-state contributions to the asymmetry, and further do not include Coulomb distortions~\cite{Gorch2}.
In addition, no corrections to Q were made to take into account the Coulomb field for the heavier nuclei.

The model calculations predict an essentially linear dependence of $B_n$ on $Q$. Such a dependence was confirmed  previously for $^{12}$C by the Mainz A1 collaboration~\cite{Esser:2018vdp}.
It is  worth noting that the far-forward angle Q$_{\rm weak}$ data on both $^{12}$C  and $^{27}$Al, as well as the far-forward angle PREX datum all lie near the upper bound of the calculations, whereas the less forward-angle Mainz $^{12}$C  and $^{28}$Si data lie near the lower bound of the predictions. Note, in particular, the significant tension between the earlier Mainz $^{12}$C datum near $Q=0.15$ GeV  and the present $^{12}$C result at a similar value of $Q$. The calculation~\cite{Esser:2018vdp} done at the 0.57 GeV appropriate for the Mainz data is almost identical to those shown in the $^{12}$C  and $^{27}$Al figures here at 1.1 GeV, an indication of the expected energy insensitivity of the calculation. Energy insensitivity in experimental results has also been observed~\cite{Androic:2020rkw} at higher energies, where far-forward $^1$H $B_n$ data at 3 GeV \cite{Armstrong:2007vm, Abrahamyan:2012cg} were shown to be consistent with 1 GeV data ~\cite{Androic:2020rkw} within  uncertainties.
Finally, it is worth noting that the new Q$_{\rm weak}$ $^{12}$C result, which required corrections for several non-elastic physics processes described in Sec.~\ref{Sec:backgrounds}, is in good agreement with the precise PREX $^{12}$C datum at the same energy (which did not require such corrections),  assuming only that the  predicted (and experimentally established~\cite{Esser:2018vdp}) $Q$-scaling is correct. However it is clear from Fig.~\ref{fig:carbon_result} when comparing the Mainz A1 datum and Q$_{\rm weak}$ datum at similar $Q \approx 0.15$ GeV but different angles and energies, that $B_n$ can also depend on $E$ or $\theta$ (only one of these is independent at fixed $Q$). Furthermore, it is also clear that the present calculations do not reproduce this additional $E/\theta$ dependence.

\subsection{\label{sec:BnvsA}Dependence of $Q$-scaled BNSSA on Mass Number $A$}

The paucity of BNSSA measurements means that it is important to compare new results to what little is already available in the literature, in the hope of shedding light on commonalities and gaining insight into the underlying physics. One way to make a comparison that uses Eq.~\ref{eq:scaling} as a foundation is to scale all the existing data to a common $Q$, and plot the results against the relevant mass number $A$. For this comparison each BNSSA result at $Q=Q_i$ was scaled to  the Q$_{\rm weak}$ average $Q=0.157$ GeV, {\em i.e.} scaled by $0.157/Q_i$. The scaling expectation shown in Eq.~\ref{eq:scaling} used  the $A$ and $Z$ for every nucleus, and the common factor $\widehat{B}_n = -30$ ppm/GeV. 

\begin{figure}[!hhtbp]
\includegraphics[width=0.75\columnwidth]{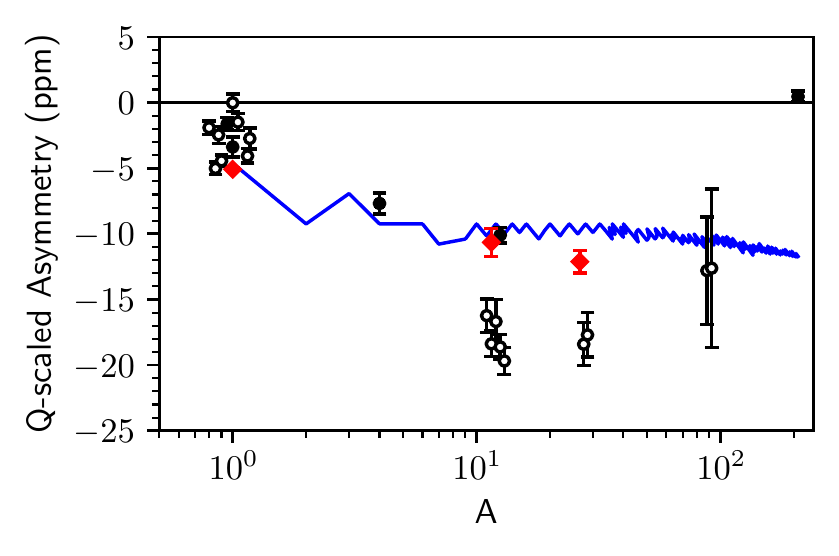}
\caption{ \label{fig:BnvsA}  All transverse asymmetries from $^1$H to $^{208}$Pb  scaled linearly in $Q$ to the average $Q=0.157$ of the Q$_{\rm weak}$ experiment  are plotted vs. atomic mass number $A$. The Q$_{\rm weak}$ data on $^1$H \cite{Androic:2018kni}, and the two new Q$_{\rm weak}$ results presented in this article on $^{12}$C and $^{27}$Al are denoted by red diamonds. Data from other experiments are represented by circles. Note that some of the data in the plot are shifted slightly in $A$ for clarity where they would otherwise overlap. Open symbols denote less far-forward angle data than denoted by solid symbols, which are generally $\theta < 10^\circ$. More backward-angle ($\theta > 50^\circ$) results are not shown, nor are results for quasi-elastic scattering on the deuteron. The blue curve represents the 
$A/Z$ dependence (Eq.~\ref{eq:scaling}) as proposed in \cite{Abrahamyan:2012cg}, with $Q=0.157$ and $\widehat{B}_n = -30$ ppm/GeV.
}
\end{figure}

The results are shown in Fig.~\ref{fig:BnvsA}. The agreement of the available data with the naive expectation of Eq.~\ref{eq:scaling} is reasonably good for the subset of the data at far-forward angles, with the single exception of the PREX $^{208}$Pb datum. This outlier has been a puzzle since it was published~\cite{Abrahamyan:2012cg}. Some have speculated that greater Coulomb distortions in $^{208}$Pb may be the key to this puzzle~\cite{Abrahamyan:2012cg}, but to date no definitive explanation exists. 

Clearly the $\theta > 10^\circ$ data (represented in Fig.~\ref{fig:BnvsA} by open symbols) are much less well described by simple scaling. This dichotomy was discussed in the previous  Sec.~\ref{sec:Gorch}, and has also been observed and discussed  in the recent Q$_{\rm weak}$ publication for $B_n$ on  $^1$H~\cite{Androic:2020rkw}. 

\subsection{\label{sec:Alldata}Q-dependence of BNSSA data scaled by $Z/A$}

\begin{figure}[h!htbp]
\includegraphics[width=1.0\columnwidth]{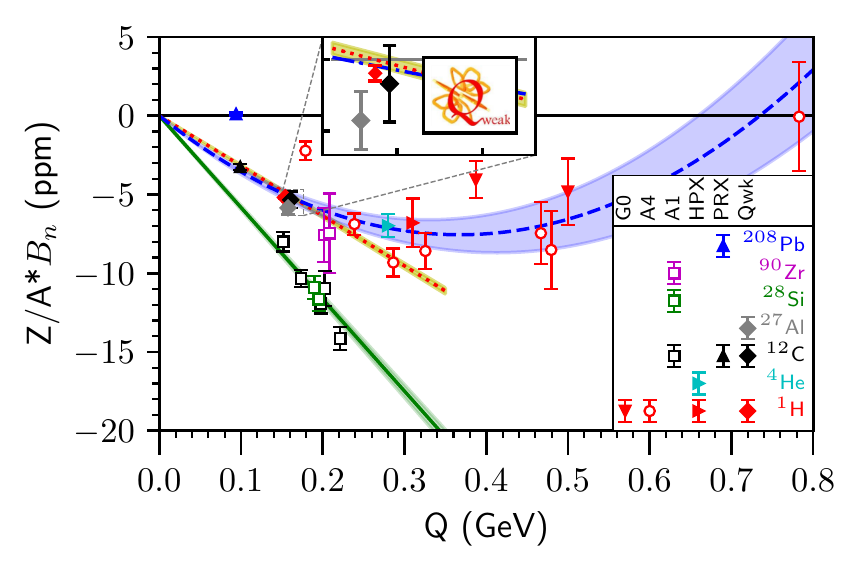}
\caption{World asymmetry data $B_n$ scaled by $Z/A$ to reduce the dependence on the target nucleus, are plotted against $Q$. Each experiment is denoted by a different symbol, each nucleus by a different color, and far-forward angle $(\theta <10^\circ )$ results (solid symbols) are differentiated from larger-angle data (open symbols). The experiments are Q$_{\rm weak}$ (\cite{Androic:2020rkw} and this experiment: {\Large $\diamond$}), HAPPEX~\cite{Abrahamyan:2012cg}:
{\large $\triangleright $},
G0~\cite{Armstrong:2007vm}: {\large $\triangledown $}, Mainz  A4~\cite{Gou:2020viq,Maas:2004pd}: {\Large $\circ$}, Mainz  A1 \cite{Esser:2018vdp,Esser:2020vjb}: $\square $, and PREX~\cite{Abrahamyan:2012cg}: {\large $\triangle $}. The target nuclei are {\color{red}\bf{$^1$H}} (red symbols), {\color{cyan}\bf{$^{4}$He}} (cyan symbol), {\color{black}\bf{$^{12}$C}} (black symbols), {\color{gray}\bf{$^{27}$Al}} (grey symbol), {\color{green}\bf{$^{28}$Si}} (green symbols), {\color{purple}\bf{$^{90}$Zr}} (magenta symbols), and {\color{blue}\bf{$^{208}$Pb}} (blue symbol). The experiments and the target nuclei are indicated in the legend.  Vertical error bars represent statistical and systematic uncertainties in quadrature. Two distinct slopes $\widehat{B}_n$ are fit $(B_n(Q=0)\equiv 0)$, as well as a fit with a quadratic term.    1-$\sigma$ fit uncertainties are denoted by the bands. The blue dashed curve (with blue band) includes a quadratic term in a fit to the Group 1 data (all $A=1$ data as well as $A>1$ far-forward angle data, but excluding the outlier $^{208}$Pb datum). The fit to the Group 1a data (red dotted line with yellow band) is the subset of the Group 1 data out to $Q<0.35$ GeV, as in \cite{Abrahamyan:2012cg}.
The steeper green dashed line (with green band) fits the slope of the Group 2 ($A>1$,  $\theta >10^\circ $) data,  which includes the $^{90}$Zr results. 
Note that some of the data in the plot (and the inset) are shifted slightly in $Q$ for clarity where they would otherwise overlap. 
\label{fig:BnvsQ_e}
}
\end{figure}

In this section, the global BNSSA data are scaled by $Z/A$ and plotted against $Q$ in Fig.~\ref{fig:BnvsQ_e}. According to Eq.~\ref{eq:scaling}, scaling each $B_n$ result by $Z/A$ should remove the nuclear dependence. 
Plotting the scaled $B_n$ against $Q$ makes it possible to empirically determine $\widehat{B}_n$  using Eq.~\ref{eq:scaling} by fitting the slopes, as long as the scaled data can be fit by a straight line.
The intercept of the fit is taken to be zero, as it is in the optical model calculation~\cite{Gorchtein:2008dy}.

\begin{table*}[!htb] \caption{Global dataset used in the figures. Some entries had to be calculated from the information provided in the  references.}
 \label{tab:Globaldata}
    \centering
    \begin{tabular}{rcccccccc}
\hline 
\rule{0pt}{3ex}	&		&	$\theta$(lab)	&	$E$(lab) 	&	$Q$	 	&	$B_n$ 	&	$\Delta B_n$ &	Fitting &		\\
	Expt	&	A	&	(deg)	&	 (GeV)	&	 (GeV)	&	 	 (ppm)	&	 (ppm)	& Group &	Ref	\\
\hline
A4	&\rule{0pt}{3ex}$^1$H	&	33.9	&	0.3151	&	0.179	&	-2.220	&	0.587	& 1,1a &	\cite{Gou:2020viq}	\\
A4	&	$^1$H	&	34.1	&	0.5102	&	0.286	&	-9.320	&	0.884	& 1,1a &	\cite{Gou:2020viq}	\\
A4	&	$^1$H	&	34.1	&	0.8552	&	0.467	&	-7.460	&	1.973	& 1 &	\cite{Gou:2020viq}	\\
A4	&	$^1$H	&	34.3	&	0.4202	&	0.239	&	-6.880	&	0.676	& 1,1a &	\cite{Gou:2020viq}	\\
A4	&	$^1$H	&	34.1	&	1.5084	&	0.783	&	-0.060	&	3.459	& 1 &	\cite{Gou:2020viq}	\\
A4	&	$^1$H	&	35.0	&	0.5693	&	0.326	&	-8.590	&	1.164	& 1,1a &	\cite{Maas:2004pd}	\\
A4	&	$^1$H	&	35.3	&	0.8552	&	0.480	&	-8.520	&	2.468	& 1 &	\cite{Maas:2004pd}	\\
G0	&	$^1$H	&	7.5	&	3.0310	&	0.387	&	-4.060	&	1.173	& 1 &	\cite{Armstrong:2007vm}	\\
G0	&	$^1$H	&	9.6	&	3.0310	&	0.500	&	-4.820	&	2.111	& 1 &	\cite{Armstrong:2007vm}	\\
Qweak	&	$^1$H	&	7.9	&	1.1490	&	0.157	&	-5.194	&	0.106	& 1,1a &	\cite{Androic:2020rkw}	\\
HAPPEX	&	$^1$H	&	6.0	&	3.0260	&	0.310	&	-6.800	&	1.540	& 1,1a &	\cite{Abrahamyan:2012cg}	\\
HAPPEX	&	$^4$He	&	6.0	&	2.7500	&	0.280	&	-13.970	&	1.450	& 1,1a &	\cite{Abrahamyan:2012cg}	\\
A1	&	$^{12}$C	&	15.1	&	0.5700	&	0.152	&	-15.984	&	1.252	& 2 &	\cite{Esser:2018vdp}	\\
A1	&	$^{12}$C	&	17.7	&	0.5700	&	0.173	&	-20.672	&	1.106	& 2 &	\cite{Esser:2018vdp}	\\
A1	&	$^{12}$C	&	20.6	&	0.5700	&	0.202	&	-21.933	&	2.219	& 2 &	\cite{Esser:2018vdp}	\\
A1	&	$^{12}$C	&	23.5	&	0.5700	&	0.197	&	-23.877	&	1.225	& 2 &	\cite{Esser:2018vdp}	\\
A1	&	$^{12}$C    &	25.9	&	0.5700	&	0.221	&	-28.296	&	1.480	& 2 &	\cite{Esser:2018vdp}	\\
PREX	&	$^{12}$C	&	5.0	&	1.0630	&	0.099	&	-6.490	&	0.380	& 1,1a &	\cite{Abrahamyan:2012cg}	\\
Qweak	&	$^{12}$C	&	7.9	&	1.1580	&	0.159	&	-10.680	&	1.065	& 1,1a &	-	\\
Qweak	&	$^{27}$Al	&	7.9	&	1.1580	&	0.154	&	-12.160	&	0.849	& 1,1a &	-	\\
A1	&	$^{28}$Si	&	19.4	&	0.5700	&	0.190	&	-21.807	&	1.480	& 2 &	\cite{Esser:2020vjb}	\\
A1	&	$^{28}$Si	&	23.5	&	0.5700	&	0.195	&	-23.302	&	1.470	& 2 &	\cite{Esser:2020vjb}	\\
A1	&	$^{90}$Zr	&	20.7	&	0.5700	&	0.205	&	-16.787	&	5.688	& 2 &	\cite{Esser:2020vjb}	\\
A1	&	$^{90}$Zr	&	23.5	&	0.5700	&	0.205	&	-17.033	&	3.848	& 2 &	\cite{Esser:2020vjb}	\\
PREX	&	$^{208}$Pb	&	5.0	&	1.0630	&	0.094	&	0.280	&	0.250	& $-$ &	\cite{Abrahamyan:2012cg}	\\
   \hline 
    \end{tabular}
\end{table*}

In Fig.~\ref{fig:BnvsQ_e} colors are used to distinguish each nucleus. Different symbols are used to distinguish each experiment. Closed or open symbols distinguish far-forward angle $(\theta \lesssim 10^\circ)$ data from larger-angle data, respectively. The 
Q$_{\rm weak}$ datum on $^1$H~\cite{Androic:2020rkw}
 as well as the new results reported here for $^{12}$C and $^{27}$Al are highlighted in the inset.
We note the remarkable fact apparent in the inset that the  factor of $\approx 2$ difference between the Q$_{\rm weak}$  $^1$H $B_n$ result and the Q$_{\rm weak}$ $B_n$ results for $^{12}$C and $^{27}$Al is almost completely eliminated by the 
$Z/A$ scaling. With this scaling, these three nuclei, all at the same kinematics, are roughly consistent with one another.

There is a lot of information to unpack in Fig.~\ref{fig:BnvsQ_e}. It is immediately clear that these data cannot be represented by a single fit.  So in order to facilitate empirical fits to these data and extract slopes $\widehat{B}_n$ using Eq.~\ref{eq:scaling}, the global BNSSA data are further separated into three groups: 
Group 1 represents all $^1$H data at any angle, as well as $\theta \lesssim 10^\circ$ far-forward angle data on any nucleus. All such BNSSA data with $0<Q<0.8$ are included in Group 1.  Group 1a is the subset of Group 1 with   $0<Q<0.35$ GeV, corresponding to the more restricted $Q$-range studied in Ref.~\cite{Abrahamyan:2012cg}. 
Group 2 contains $A>1$, $(\theta > 10^\circ)$ data, which consist of the Mainz $^{12}$C~\cite{Esser:2018vdp}, and the Mainz $^{28}$Si and $^{90}$Zr data~\cite{Esser:2020vjb}. These data clearly have a steeper slope than those in Group 1 (or its subset Group 1a) and thus require a separate fit. 
The $^{208}$Pb outlier datum~\cite{Abrahamyan:2012cg}  does not  fit into any group, is not included in any of the fits discussed here, and no attempt to assign a slope to this datum is made. 

The fit to the Group 1 data obviously requires a non-linear component in order to describe the data at higher $Q$. A similar deviation from linear scaling at higher $Q$ was predicted in the optical theorem approach by Afanasev and Merenkov~\cite{Afanasev:2004pu} for $B_n$ for the proton. Since the focus here is an empirical/phenomenological characterization of the global $B_n$ data, a quadratic term is simply added to the fit of the Group 1 data, as shown in Table~\ref{tab:Bnfits}. To be clear, this fit 
returns the linear slope $\widehat{B}_n$ and quadratic term $\beta$ from 
$Z/A \; B_n = \widehat{B}_n Q + \beta Q^2$. Fits to Groups 1a and 2 drop the quadratic term. The Group 1 and 1a fits are tightly constrained by the unusually good precision of both the
Q$_{\rm weak}$ $^1$H datum~\cite{Androic:2020rkw} as well as the PREX $^{12}$C datum~\cite{Abrahamyan:2012cg}. The datum contributing most to the $\chi^2/\rm{dof}$ of those two  fits  is the lowest-$Q$ datum from the Mainz A4 $^1$H results~\cite{Gou:2020viq}.

In order to compare more directly with Ref.~\cite{Abrahamyan:2012cg}, and to avoid the non-linear behaviour shown by the higher-$Q$ data, Group 1a is the subset of Group 1 with $Q<0.35$ GeV. No quadratic term is used in the Group 1a fit result shown in Table~\ref{tab:Bnfits}. 

The $^{27}$Al and $^{12}$C  calculations shown in Figs.~\ref{fig:aluminum_result} and \ref{fig:carbon_result}, times $Z/A$,  both have an effective slope $\widehat{B}_n=-44.0 \pm^{11.0}_{12.7}$ ppm/GeV. This result is consistent with all the empirical fits found for the Groups 1 \& 1a data.

The Group 2 data have twice  the slope of the other fits, as shown in Table~\ref{tab:Bnfits}. This group includes the  larger-angle Mainz A1 data on  $^{12}$C~\cite{Esser:2018vdp}, as well as their  $^{28}$Si and $^{90}$Zr results~\cite{Esser:2020vjb}. The  $^{90}$Zr data belong in this group because they were part of the same experiment as $^{28}$Si and had similar kinematics. They appear to be 
more consistent with the fit that has the shallower slope, but because they are in the larger angle $(\theta > 10^\circ)$, $A>1$ group they are included in the fit to those data. The lower bound of the slope associated with the theoretical calculations in Figs.~\ref{fig:aluminum_result} and \ref{fig:carbon_result} is consistent with the empirical fits found for the Group 2 data.

\begin{table*}[!htb] \caption{Fit results. }
 \label{tab:Bnfits}
    \centering
    \begin{tabular}{ccccc}
\hline 
\vspace{0.8cm}
   \rule{0pt}{3ex} & Linear $(\widehat{B}_n)$  & Quadratic ($\beta$) &  &  \\ 
    Group & (ppm/GeV)  & (ppm/GeV$^2$) & \# data & $\chi^2/{\rm dof}$  \\ \hline
    1  &    $-41.1 \pm 1.1$ & $56.0 \pm 4.8$ & 15 & 4.4 \\
    1a  &    $-31.8 \pm 0.5$ & $-$ & 10 & 6.4 \\
    2  &    $-58.3 \pm 1.4 $ & $-$ & 9 & 2.0 \\
   \hline 
    \end{tabular}
\end{table*}

\clearpage
\subsection{\label{sec:Bnhat} $Q$-dependence of $\widehat{B}_n$}

The previous section examined consistencies in the data apparent  once the nuclear dependence was removed via $Z/A$ scaling. The Qweak results on $^1$H, $^{12}$C, and $^{27}$Al were particularly revealing, as those results are at the same kinematics and were seen to be consistent after scaling. 

\begin{figure}[!htbp]
\includegraphics[width=0.75\columnwidth]{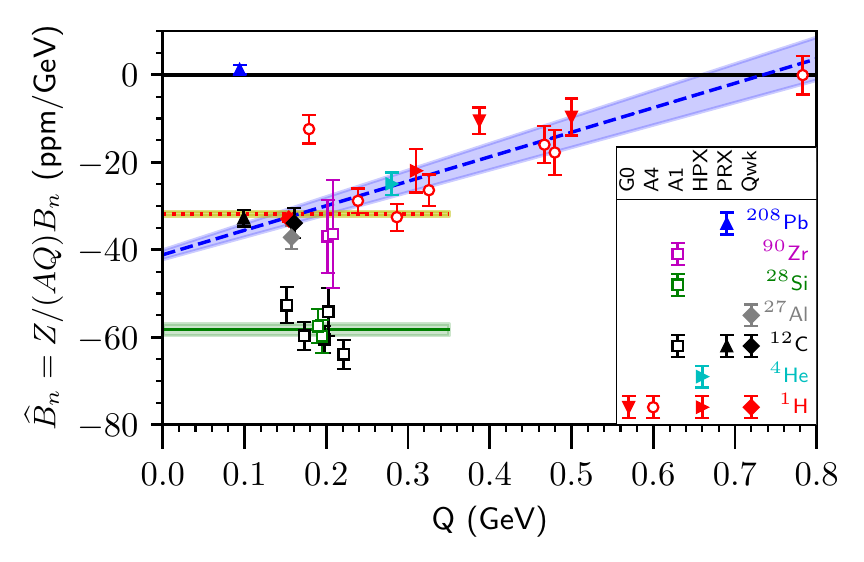}
\caption{World transverse asymmetry data $B_n$ are scaled by the factor $Z/(AQ)$ 
and plotted  against Q (in GeV). 
Symbols and colors are as in Fig.~\ref{fig:BnvsQ_e}. The dotted red and solid green lines represent fits to the intercepts  of the Group 1a \& 2 data, respectively, and thus correspond to the slopes $\widehat{B}_n$ in Fig.~\ref{fig:BnvsQ_e} and Table~\ref{tab:Bnfits}.
The blue dashed line is a linear fit to all the Group 1 data, corresponding to the quadratic fit in Fig.~\ref{fig:BnvsQ_e} and Table~\ref{tab:Bnfits}. Uncertainties in the fits are denoted by the bands in the figure.
\label{fig:Bnhat} }
\end{figure}

In this section we remove the explicit $Q$-dependence as well as the nuclear dependence, and plot $\widehat{B}_n = \frac{Z}{A}\frac{1}{Q}B_n$ versus $Q$.
The expectation from Eq.~\ref{eq:scaling} is that such a plot
would consist of data that could be represented with a flat horizontal line, because $\widehat{B}_n$ is assumed to be a constant.  This is shown in Fig.~\ref{fig:Bnhat}, where the same categories are used to group the data for fitting as were used in Fig.~\ref{fig:BnvsQ_e}. Fitting the Group 1a and 2 data in Fig.~\ref{fig:Bnhat} with the assumption that they are flat horizontal lines results in $\widehat{B}_n$ intercept values and uncertainties identical to those obtained in the previous section~\ref{sec:Alldata} and tabulated in Table~\ref{tab:Bnfits} for the slopes $\widehat{B}_n$ found in those fits.

However, it is clear that the  higher $Q$ Group 1  data in Fig.~\ref{fig:Bnhat} have a residual $Q$-dependence, which we empirically model as linear: 
$\widehat{B}_n = \widehat{B}_n^{\, 0}(Q=0) + Q\: \widehat{B}^{\, Q}_n$. 
The first term is the intercept. The second term is responsible for the residual $Q$-dependence seen in Fig.~\ref{fig:Bnhat}, and the quadratic behaviour seen in Fig.~\ref{fig:BnvsQ_e}.
In response, the Group 1 data (up to 0.8 GeV) were fit to  determine an intercept  as well as a slope. The numerical values and uncertainties returned from the fit  intercept ($\widehat{B}_n$) and slope are identical to the values found in the previous section (Sec.~\ref{sec:Alldata}) for the linear slope ($\widehat{B}_n$) and quadratic terms  respectively, for the Group 1 fit shown in Table~\ref{tab:Bnfits}.
The Group 2 data in Fig.~\ref{fig:Bnhat} do not have a sufficient range in $Q$ to justify fitting a slope to them. 

\section{Conclusions}
The beam-normal single-spin asymmetry $B_n$ has been measured at 
forward-angle kinematics 
for $^{12}$C and $^{27}$Al. At small scattering angles, a model for $B_n$ based on the optical theorem ~\cite{Gorchtein:2008dy} is expected to be valid. This model is able to reproduce both 
of the measurements  reported here within the uncertainty of the calculation. 
The new Q$_{\rm weak}$ $^{12}$C result together with the PREX datum at a lower $Q$ but similar scattering angle are in excellent agreement with the  predicted $Q$-dependence. Comparing with earlier data on $^{12}$C obtained 
at larger laboratory scattering angles suggests that for those kinematics the model's reliance on taking  the far-forward approximation may be reaching its limit of applicability. Similar conclusions are drawn from a comparison of the 
new Q$_{\rm weak}$  $^{27}$Al result with previous results for $^{28}$Si. 
These comparisons also suggest that modest contributions from nuclear excited states can be successfully accounted for in measurements of $B_n$.

A global analysis of world $B_n$ data at forward angles supports these conclusions: a simple linear scaling $Z/(AQ)$ works well for most data at far forward-angles for $Q<0.35$~GeV. Data at larger angles follow a steeper $Q$-dependence. For $Q>0.35$ the dependence on $Q$ is clearly non-linear, and can be empirically modelled by a quadratic dependence.  If  further divided by $Q$, this quadratic dependence appears linear out to 0.8 GeV for all the world's $^1$H data as well as far-forward angle ($\theta < 10^\circ$) data on any nucleus. The significant exception to these trends is the case of $^{208}$Pb, whose unexpectedly small BNSSA remains unexplained. 
Data on $B_n$ have recently been obtained by the PREX-2/CREX collaborations~\cite{McNulty:2020,Richards:2020} for two isotopes of Ca, as well as new measurements of $^{12}$C and $^{208}$Pb, which may shed additional light  on the $^{208}$Pb puzzle.
Finally, as this paper was being completed, a preprint appeared by 
Koshchii, Gorchtein, Roca-Maza, and Spiesberger ~\cite{Koshchii:2021mqq} in which $B_n$ for selected nuclei was calculated with inclusion of both hard two-photon exchange and Coulomb distortions. That model's predictions of $B_n$ for $^{12}$C and $^{27}$Al at the present kinematics are lower in magnitude than the calculations displayed in Figs. \ref{fig:aluminum_result} and \ref{fig:carbon_result}, but still consistent with our data.

\begin{acknowledgments}
We thank the staff of Jefferson Lab, in particular the accelerator operations staff, 
the radiation control staff, as well as the Hall C technical staff for their help and support. We are also grateful for the contributions of our undergraduate students. We thank TRIUMF for its contributions to the development of the spectrometer and integrated electronics, and BATES for its contributions to the spectrometer and Compton polarimeter. 
We are indebted to C. Horowitz and Z. Lin for their cross section calculations.
We thank M. Gorchtein for helpful discussions and unpublished calculations. This material is based upon work supported by the U.S. Department of Energy, Office of Science, Office of Nuclear Physics under contract DE-AC05-06OR23177.
Construction and operating funding for the experiment was provided through the DOE, the Natural Sciences and Engineering Research Council of Canada (NSERC), the Canadian Foundation for Innovation (CFI), and the National Science Foundation (NSF) with university matching contributions from William \& Mary, Virginia Tech, George Washington University and Louisiana Tech University. 
\end{acknowledgments}


\bibliography{qweak_aluminum_transverse}

\begin{thebibliography}{87}%
\makeatletter
\providecommand \@ifxundefined [1]{%
 \@ifx{#1\undefined}
}%
\providecommand \@ifnum [1]{%
 \ifnum #1\expandafter \@firstoftwo
 \else \expandafter \@secondoftwo
 \fi
}%
\providecommand \@ifx [1]{%
 \ifx #1\expandafter \@firstoftwo
 \else \expandafter \@secondoftwo
 \fi
}%
\providecommand \natexlab [1]{#1}%
\providecommand \enquote  [1]{``#1''}%
\providecommand \bibnamefont  [1]{#1}%
\providecommand \bibfnamefont [1]{#1}%
\providecommand \citenamefont [1]{#1}%
\providecommand \href@noop [0]{\@secondoftwo}%
\providecommand \href [0]{\begingroup \@sanitize@url \@href}%
\providecommand \@href[1]{\@@startlink{#1}\@@href}%
\providecommand \@@href[1]{\endgroup#1\@@endlink}%
\providecommand \@sanitize@url [0]{\catcode `\\12\catcode `\$12\catcode
  `\&12\catcode `\#12\catcode `\^12\catcode `\_12\catcode `\%12\relax}%
\providecommand \@@startlink[1]{}%
\providecommand \@@endlink[0]{}%
\providecommand \url  [0]{\begingroup\@sanitize@url \@url }%
\providecommand \@url [1]{\endgroup\@href {#1}{\urlprefix }}%
\providecommand \urlprefix  [0]{URL }%
\providecommand \Eprint [0]{\href }%
\providecommand \doibase [0]{http://dx.doi.org/}%
\providecommand \selectlanguage [0]{\@gobble}%
\providecommand \bibinfo  [0]{\@secondoftwo}%
\providecommand \bibfield  [0]{\@secondoftwo}%
\providecommand \translation [1]{[#1]}%
\providecommand \BibitemOpen [0]{}%
\providecommand \bibitemStop [0]{}%
\providecommand \bibitemNoStop [0]{.\EOS\space}%
\providecommand \EOS [0]{\spacefactor3000\relax}%
\providecommand \BibitemShut  [1]{\csname bibitem#1\endcsname}%
\let\auto@bib@innerbib\@empty
\bibitem [{\citenamefont {Walecka}(2005)}]{Walecka:2001gs}%
  \BibitemOpen
  \bibfield  {author} {\bibinfo {author} {\bibfnamefont {J.~D.}\ \bibnamefont
  {Walecka}},\ }\href@noop {} {\emph {\bibinfo {title} {{Electron scattering
  for nuclear and nucleon structure}}}}\ (\bibinfo  {publisher} {Cambridge
  University Press},\ \bibinfo {year} {2005})\BibitemShut {NoStop}%
\bibitem [{\citenamefont {Afanasev}\ \emph {et~al.}(2017)\citenamefont
  {Afanasev}, \citenamefont {Blunden}, \citenamefont {Hasell},\ and\
  \citenamefont {Raue}}]{Afanasev:2017gsk}%
  \BibitemOpen
  \bibfield  {author} {\bibinfo {author} {\bibfnamefont {A.}~\bibnamefont
  {Afanasev}}, \bibinfo {author} {\bibfnamefont {P.~G.}\ \bibnamefont
  {Blunden}}, \bibinfo {author} {\bibfnamefont {D.}~\bibnamefont {Hasell}}, \
  and\ \bibinfo {author} {\bibfnamefont {B.~A.}\ \bibnamefont {Raue}},\ }\href
  {\doibase 10.1016/j.ppnp.2017.03.004} {\bibfield  {journal} {\bibinfo
  {journal} {Prog. Part. Nucl. Phys.}\ }\textbf {\bibinfo {volume} {95}},\
  \bibinfo {pages} {245} (\bibinfo {year} {2017})}\BibitemShut {NoStop}%
\bibitem [{\citenamefont {Guichon}\ and\ \citenamefont
  {Vanderhaeghen}(2003)}]{Guichon:2003qm}%
  \BibitemOpen
  \bibfield  {author} {\bibinfo {author} {\bibfnamefont {P.~A.~M.}\
  \bibnamefont {Guichon}}\ and\ \bibinfo {author} {\bibfnamefont
  {M.}~\bibnamefont {Vanderhaeghen}},\ }\href {\doibase
  10.1103/PhysRevLett.91.142303} {\bibfield  {journal} {\bibinfo  {journal}
  {Phys. Rev. Lett.}\ }\textbf {\bibinfo {volume} {91}},\ \bibinfo {pages}
  {142303} (\bibinfo {year} {2003})}\BibitemShut {NoStop}%
\bibitem [{\citenamefont {Blunden}\ \emph {et~al.}(2003)\citenamefont
  {Blunden}, \citenamefont {Melnitchouk},\ and\ \citenamefont
  {Tjon}}]{Blunden:2003sp}%
  \BibitemOpen
  \bibfield  {author} {\bibinfo {author} {\bibfnamefont {P.~G.}\ \bibnamefont
  {Blunden}}, \bibinfo {author} {\bibfnamefont {W.}~\bibnamefont
  {Melnitchouk}}, \ and\ \bibinfo {author} {\bibfnamefont {J.~A.}\ \bibnamefont
  {Tjon}},\ }\href {\doibase 10.1103/PhysRevLett.91.142304} {\bibfield
  {journal} {\bibinfo  {journal} {Phys. Rev. Lett.}\ }\textbf {\bibinfo
  {volume} {91}},\ \bibinfo {pages} {142304} (\bibinfo {year}
  {2003})}\BibitemShut {NoStop}%
\bibitem [{\citenamefont {Gorchtein}\ and\ \citenamefont
  {Horowitz}(2009)}]{Gorchtein:2008px}%
  \BibitemOpen
  \bibfield  {author} {\bibinfo {author} {\bibfnamefont {M.}~\bibnamefont
  {Gorchtein}}\ and\ \bibinfo {author} {\bibfnamefont {C.~J.}\ \bibnamefont
  {Horowitz}},\ }\href {\doibase 10.1103/PhysRevLett.102.091806} {\bibfield
  {journal} {\bibinfo  {journal} {Phys. Rev. Lett.}\ }\textbf {\bibinfo
  {volume} {102}},\ \bibinfo {pages} {091806} (\bibinfo {year}
  {2009})}\BibitemShut {NoStop}%
\bibitem [{\citenamefont {Sibirtsev}\ \emph {et~al.}(2010)\citenamefont
  {Sibirtsev}, \citenamefont {Blunden}, \citenamefont {Melnitchouk},\ and\
  \citenamefont {Thomas}}]{Sibirtsev:2010zg}%
  \BibitemOpen
  \bibfield  {author} {\bibinfo {author} {\bibfnamefont {A.}~\bibnamefont
  {Sibirtsev}}, \bibinfo {author} {\bibfnamefont {P.~G.}\ \bibnamefont
  {Blunden}}, \bibinfo {author} {\bibfnamefont {W.}~\bibnamefont
  {Melnitchouk}}, \ and\ \bibinfo {author} {\bibfnamefont {A.~W.}\ \bibnamefont
  {Thomas}},\ }\href {\doibase 10.1103/PhysRevD.82.013011} {\bibfield
  {journal} {\bibinfo  {journal} {Phys. Rev.}\ }\textbf {\bibinfo {volume}
  {D82}},\ \bibinfo {pages} {013011} (\bibinfo {year} {2010})}\BibitemShut
  {NoStop}%
\bibitem [{\citenamefont {Rislow}\ and\ \citenamefont
  {Carlson}(2011)}]{Rislow:2010vi}%
  \BibitemOpen
  \bibfield  {author} {\bibinfo {author} {\bibfnamefont {B.~C.}\ \bibnamefont
  {Rislow}}\ and\ \bibinfo {author} {\bibfnamefont {C.~E.}\ \bibnamefont
  {Carlson}},\ }\href {\doibase 10.1103/PhysRevD.83.113007} {\bibfield
  {journal} {\bibinfo  {journal} {Phys. Rev.}\ }\textbf {\bibinfo {volume}
  {D83}},\ \bibinfo {pages} {113007} (\bibinfo {year} {2011})}\BibitemShut
  {NoStop}%
\bibitem [{\citenamefont {Rislow}\ and\ \citenamefont
  {Carlson}(2013)}]{Rislow:2013vta}%
  \BibitemOpen
  \bibfield  {author} {\bibinfo {author} {\bibfnamefont {B.~C.}\ \bibnamefont
  {Rislow}}\ and\ \bibinfo {author} {\bibfnamefont {C.~E.}\ \bibnamefont
  {Carlson}},\ }\href {\doibase 10.1103/PhysRevD.88.013018} {\bibfield
  {journal} {\bibinfo  {journal} {Phys. Rev.}\ }\textbf {\bibinfo {volume}
  {D88}},\ \bibinfo {pages} {013018} (\bibinfo {year} {2013})}\BibitemShut
  {NoStop}%
\bibitem [{\citenamefont {Gorchtein}\ \emph {et~al.}(2011)\citenamefont
  {Gorchtein}, \citenamefont {Horowitz},\ and\ \citenamefont
  {Ramsey-Musolf}}]{Gorchtein:2011mz}%
  \BibitemOpen
  \bibfield  {author} {\bibinfo {author} {\bibfnamefont {M.}~\bibnamefont
  {Gorchtein}}, \bibinfo {author} {\bibfnamefont {C.~J.}\ \bibnamefont
  {Horowitz}}, \ and\ \bibinfo {author} {\bibfnamefont {M.~J.}\ \bibnamefont
  {Ramsey-Musolf}},\ }\href {\doibase 10.1103/PhysRevC.84.015502} {\bibfield
  {journal} {\bibinfo  {journal} {Phys. Rev.}\ }\textbf {\bibinfo {volume}
  {C84}},\ \bibinfo {pages} {015502} (\bibinfo {year} {2011})}\BibitemShut
  {NoStop}%
\bibitem [{\citenamefont {Blunden}\ \emph {et~al.}(2011)\citenamefont
  {Blunden}, \citenamefont {Melnitchouk},\ and\ \citenamefont
  {Thomas}}]{Blunden:2011rd}%
  \BibitemOpen
  \bibfield  {author} {\bibinfo {author} {\bibfnamefont {P.~G.}\ \bibnamefont
  {Blunden}}, \bibinfo {author} {\bibfnamefont {W.}~\bibnamefont
  {Melnitchouk}}, \ and\ \bibinfo {author} {\bibfnamefont {A.~W.}\ \bibnamefont
  {Thomas}},\ }\href {\doibase 10.1103/PhysRevLett.107.081801} {\bibfield
  {journal} {\bibinfo  {journal} {Phys. Rev. Lett.}\ }\textbf {\bibinfo
  {volume} {107}},\ \bibinfo {pages} {081801} (\bibinfo {year}
  {2011})}\BibitemShut {NoStop}%
\bibitem [{\citenamefont {Hall}\ \emph {et~al.}(2013)\citenamefont {Hall},
  \citenamefont {Blunden}, \citenamefont {Melnitchouk}, \citenamefont
  {Thomas},\ and\ \citenamefont {Young}}]{Hall:2013hta}%
  \BibitemOpen
  \bibfield  {author} {\bibinfo {author} {\bibfnamefont {N.~L.}\ \bibnamefont
  {Hall}}, \bibinfo {author} {\bibfnamefont {P.~G.}\ \bibnamefont {Blunden}},
  \bibinfo {author} {\bibfnamefont {W.}~\bibnamefont {Melnitchouk}}, \bibinfo
  {author} {\bibfnamefont {A.~W.}\ \bibnamefont {Thomas}}, \ and\ \bibinfo
  {author} {\bibfnamefont {R.~D.}\ \bibnamefont {Young}},\ }\href {\doibase
  10.1103/PhysRevD.88.013011} {\bibfield  {journal} {\bibinfo  {journal} {Phys.
  Rev.}\ }\textbf {\bibinfo {volume} {D88}},\ \bibinfo {pages} {013011}
  (\bibinfo {year} {2013})}\BibitemShut {NoStop}%
\bibitem [{\citenamefont {Hall}\ \emph {et~al.}(2016)\citenamefont {Hall},
  \citenamefont {Blunden}, \citenamefont {Melnitchouk}, \citenamefont
  {Thomas},\ and\ \citenamefont {Young}}]{Hall:2015loa}%
  \BibitemOpen
  \bibfield  {author} {\bibinfo {author} {\bibfnamefont {N.~L.}\ \bibnamefont
  {Hall}}, \bibinfo {author} {\bibfnamefont {P.~G.}\ \bibnamefont {Blunden}},
  \bibinfo {author} {\bibfnamefont {W.}~\bibnamefont {Melnitchouk}}, \bibinfo
  {author} {\bibfnamefont {A.~W.}\ \bibnamefont {Thomas}}, \ and\ \bibinfo
  {author} {\bibfnamefont {R.~D.}\ \bibnamefont {Young}},\ }\href {\doibase
  10.1016/j.physletb.2015.11.081} {\bibfield  {journal} {\bibinfo  {journal}
  {Phys. Lett.}\ }\textbf {\bibinfo {volume} {B753}},\ \bibinfo {pages} {221}
  (\bibinfo {year} {2016})}\BibitemShut {NoStop}%
\bibitem [{\citenamefont {Androi\'c}\ \emph {et~al.}(2018)\citenamefont
  {Androi\'c} \emph {et~al.}}]{Androic:2018kni}%
  \BibitemOpen
  \bibfield  {author} {\bibinfo {author} {\bibfnamefont {D.}~\bibnamefont
  {Androi\'c}} \emph {et~al.} (\bibinfo {collaboration} {Q$_{\rm weak}$
  Collaboration}),\ }\href {\doibase 10.1038/s41586-018-0096-0} {\bibfield
  {journal} {\bibinfo  {journal} {Nature}\ }\textbf {\bibinfo {volume} {557}},\
  \bibinfo {pages} {207} (\bibinfo {year} {2018})}\BibitemShut {NoStop}%
\bibitem [{\citenamefont {Androi\'c}\ \emph {et~al.}(2013)\citenamefont
  {Androi\'c} \emph {et~al.}}]{Androic:2013rhu}%
  \BibitemOpen
  \bibfield  {author} {\bibinfo {author} {\bibfnamefont {D.}~\bibnamefont
  {Androi\'c}} \emph {et~al.} (\bibinfo {collaboration} {Q$_{\rm weak}$
  Collaboration}),\ }\href {\doibase 10.1103/PhysRevLett.111.141803} {\bibfield
   {journal} {\bibinfo  {journal} {Phys. Rev. Lett.}\ }\textbf {\bibinfo
  {volume} {111}},\ \bibinfo {pages} {141803} (\bibinfo {year}
  {2013})}\BibitemShut {NoStop}%
\bibitem [{\citenamefont {Becker}\ \emph {et~al.}(2018)\citenamefont {Becker}
  \emph {et~al.}}]{Becker:2018gzk}%
  \BibitemOpen
  \bibfield  {author} {\bibinfo {author} {\bibfnamefont {D.}~\bibnamefont
  {Becker}} \emph {et~al.},\ }\href {\doibase 10.1140/epja/i2018-12611-6}
  {\bibfield  {journal} {\bibinfo  {journal} {Eur. Phys. J.}\ }\textbf
  {\bibinfo {volume} {A54}},\ \bibinfo {pages} {208} (\bibinfo {year}
  {2018})}\BibitemShut {NoStop}%
\bibitem [{\citenamefont {Seng}\ \emph {et~al.}(2018)\citenamefont {Seng},
  \citenamefont {Gorchtein}, \citenamefont {Patel},\ and\ \citenamefont
  {Ramsey-Musolf}}]{PhysRevLett.121.241804}%
  \BibitemOpen
  \bibfield  {author} {\bibinfo {author} {\bibfnamefont {C.-Y.}\ \bibnamefont
  {Seng}}, \bibinfo {author} {\bibfnamefont {M.}~\bibnamefont {Gorchtein}},
  \bibinfo {author} {\bibfnamefont {H.~H.}\ \bibnamefont {Patel}}, \ and\
  \bibinfo {author} {\bibfnamefont {M.~J.}\ \bibnamefont {Ramsey-Musolf}},\
  }\href {\doibase 10.1103/PhysRevLett.121.241804} {\bibfield  {journal}
  {\bibinfo  {journal} {Phys. Rev. Lett.}\ }\textbf {\bibinfo {volume} {121}},\
  \bibinfo {pages} {241804} (\bibinfo {year} {2018})}\BibitemShut {NoStop}%
\bibitem [{\citenamefont {Aste}\ \emph {et~al.}(2005)\citenamefont {Aste},
  \citenamefont {von Arx},\ and\ \citenamefont {Trautmann}}]{Article:Aste2005}%
  \BibitemOpen
  \bibfield  {author} {\bibinfo {author} {\bibfnamefont {A.}~\bibnamefont
  {Aste}}, \bibinfo {author} {\bibfnamefont {C.}~\bibnamefont {von Arx}}, \
  and\ \bibinfo {author} {\bibfnamefont {D.}~\bibnamefont {Trautmann}},\ }\href
  {\doibase 10.1140/epja/i2005-10169-0} {\bibfield  {journal} {\bibinfo
  {journal} {Eur. Phys. J.}\ }\textbf {\bibinfo {volume} {A26}},\ \bibinfo
  {pages} {167} (\bibinfo {year} {2005})}\BibitemShut {NoStop}%
\bibitem [{\citenamefont {Barschall}\ and\ \citenamefont
  {Haeberli}(1971)}]{osti_4726823}%
  \BibitemOpen
  \bibinfo {editor} {\bibfnamefont {H.}~\bibnamefont {Barschall}}\ and\
  \bibinfo {editor} {\bibfnamefont {W.}~\bibnamefont {Haeberli}},\ eds.,\
  \href@noop {} {\emph {\bibinfo {title} {Polarization Phenomena in Nuclear
  Reactions, Proc. Third Int. Symposium, Madison, Wisconsin}}}\ (\bibinfo
  {publisher} {U. Wisconson Press},\ \bibinfo {year} {1971})\BibitemShut
  {NoStop}%
\bibitem [{\citenamefont {De~Rujula}\ \emph {et~al.}(1971)\citenamefont
  {De~Rujula}, \citenamefont {Kaplan},\ and\ \citenamefont
  {De~Rafael}}]{DeRujula:1972te}%
  \BibitemOpen
  \bibfield  {author} {\bibinfo {author} {\bibfnamefont {A.}~\bibnamefont
  {De~Rujula}}, \bibinfo {author} {\bibfnamefont {J.}~\bibnamefont {Kaplan}}, \
  and\ \bibinfo {author} {\bibfnamefont {E.}~\bibnamefont {De~Rafael}},\ }\href
  {\doibase 10.1016/0550-3213(71)90460-3} {\bibfield  {journal} {\bibinfo
  {journal} {Nucl. Phys.}\ }\textbf {\bibinfo {volume} {B35}},\ \bibinfo
  {pages} {365} (\bibinfo {year} {1971})}\BibitemShut {NoStop}%
\bibitem [{\citenamefont {Wells}\ \emph {et~al.}(2001)\citenamefont {Wells}
  \emph {et~al.}}]{Wells:2000rx}%
  \BibitemOpen
  \bibfield  {author} {\bibinfo {author} {\bibfnamefont {S.~P.}\ \bibnamefont
  {Wells}} \emph {et~al.} (\bibinfo {collaboration} {SAMPLE Collaboration}),\
  }\href {\doibase 10.1103/PhysRevC.63.064001} {\bibfield  {journal} {\bibinfo
  {journal} {Phys. Rev.}\ }\textbf {\bibinfo {volume} {C63}},\ \bibinfo {pages}
  {064001} (\bibinfo {year} {2001})}\BibitemShut {NoStop}%
\bibitem [{\citenamefont {Grames}\ \emph {et~al.}(2020)\citenamefont {Grames}
  \emph {et~al.}}]{Grames:2020asy}%
  \BibitemOpen
  \bibfield  {author} {\bibinfo {author} {\bibfnamefont {J.~M.}\ \bibnamefont
  {Grames}} \emph {et~al.},\ }\href {\doibase 10.1103/PhysRevC.102.015501}
  {\bibfield  {journal} {\bibinfo  {journal} {Phys. Rev. C}\ }\textbf {\bibinfo
  {volume} {102}},\ \bibinfo {pages} {015501} (\bibinfo {year}
  {2020})}\BibitemShut {NoStop}%
\bibitem [{\citenamefont {Rachek}\ \emph {et~al.}(2015)\citenamefont {Rachek}
  \emph {et~al.}}]{Rachek:2014fam}%
  \BibitemOpen
  \bibfield  {author} {\bibinfo {author} {\bibfnamefont {I.~A.}\ \bibnamefont
  {Rachek}} \emph {et~al.},\ }\href {\doibase 10.1103/PhysRevLett.114.062005}
  {\bibfield  {journal} {\bibinfo  {journal} {Phys. Rev. Lett.}\ }\textbf
  {\bibinfo {volume} {114}},\ \bibinfo {pages} {062005} (\bibinfo {year}
  {2015})}\BibitemShut {NoStop}%
\bibitem [{\citenamefont {Adikaram}\ \emph {et~al.}(2015)\citenamefont
  {Adikaram} \emph {et~al.}}]{Adikaram:2014ykv}%
  \BibitemOpen
  \bibfield  {author} {\bibinfo {author} {\bibfnamefont {D.}~\bibnamefont
  {Adikaram}} \emph {et~al.} (\bibinfo {collaboration} {CLAS Collaboration}),\
  }\href {\doibase 10.1103/PhysRevLett.114.062003} {\bibfield  {journal}
  {\bibinfo  {journal} {Phys. Rev. Lett.}\ }\textbf {\bibinfo {volume} {114}},\
  \bibinfo {pages} {062003} (\bibinfo {year} {2015})}\BibitemShut {NoStop}%
\bibitem [{\citenamefont {Henderson}\ \emph {et~al.}(2017)\citenamefont
  {Henderson} \emph {et~al.}}]{Henderson:2016dea}%
  \BibitemOpen
  \bibfield  {author} {\bibinfo {author} {\bibfnamefont {B.~S.}\ \bibnamefont
  {Henderson}} \emph {et~al.} (\bibinfo {collaboration} {OLYMPUS
  Collaboration}),\ }\href {\doibase 10.1103/PhysRevLett.118.092501} {\bibfield
   {journal} {\bibinfo  {journal} {Phys. Rev. Lett.}\ }\textbf {\bibinfo
  {volume} {118}},\ \bibinfo {pages} {092501} (\bibinfo {year}
  {2017})}\BibitemShut {NoStop}%
\bibitem [{\citenamefont {Afanasev}\ and\ \citenamefont
  {Merenkov}(2004)}]{Afanasev:2004pu}%
  \BibitemOpen
  \bibfield  {author} {\bibinfo {author} {\bibfnamefont {A.~V.}\ \bibnamefont
  {Afanasev}}\ and\ \bibinfo {author} {\bibfnamefont {N.}~\bibnamefont
  {Merenkov}},\ }\href {\doibase 10.1016/j.physletb.2004.08.023} {\bibfield
  {journal} {\bibinfo  {journal} {Phys. Lett.}\ }\textbf {\bibinfo {volume}
  {B599}},\ \bibinfo {pages} {48} (\bibinfo {year} {2004})}\BibitemShut
  {NoStop}%
\bibitem [{\citenamefont {Pasquini}\ and\ \citenamefont
  {Vanderhaeghen}(2004)}]{Pasquini:2004pv}%
  \BibitemOpen
  \bibfield  {author} {\bibinfo {author} {\bibfnamefont {B.}~\bibnamefont
  {Pasquini}}\ and\ \bibinfo {author} {\bibfnamefont {M.}~\bibnamefont
  {Vanderhaeghen}},\ }\href {\doibase 10.1103/PhysRevC.70.045206} {\bibfield
  {journal} {\bibinfo  {journal} {Phys. Rev.}\ }\textbf {\bibinfo {volume}
  {C70}},\ \bibinfo {pages} {045206} (\bibinfo {year} {2004})}\BibitemShut
  {NoStop}%
\bibitem [{\citenamefont {Tomalak}\ \emph {et~al.}(2017)\citenamefont
  {Tomalak}, \citenamefont {Pasquini},\ and\ \citenamefont
  {Vanderhaeghen}}]{Tomalak:2016vbf}%
  \BibitemOpen
  \bibfield  {author} {\bibinfo {author} {\bibfnamefont {O.}~\bibnamefont
  {Tomalak}}, \bibinfo {author} {\bibfnamefont {B.}~\bibnamefont {Pasquini}}, \
  and\ \bibinfo {author} {\bibfnamefont {M.}~\bibnamefont {Vanderhaeghen}},\
  }\href {\doibase 10.1103/PhysRevD.95.096001} {\bibfield  {journal} {\bibinfo
  {journal} {Phys. Rev. D}\ }\textbf {\bibinfo {volume} {95}},\ \bibinfo
  {pages} {096001} (\bibinfo {year} {2017})}\BibitemShut {NoStop}%
\bibitem [{\citenamefont {Gorchtein}\ \emph {et~al.}(2004)\citenamefont
  {Gorchtein}, \citenamefont {Guichon},\ and\ \citenamefont
  {Vanderhaeghen}}]{Gorchtein:2004ac}%
  \BibitemOpen
  \bibfield  {author} {\bibinfo {author} {\bibfnamefont {M.}~\bibnamefont
  {Gorchtein}}, \bibinfo {author} {\bibfnamefont {P.~A.~M.}\ \bibnamefont
  {Guichon}}, \ and\ \bibinfo {author} {\bibfnamefont {M.}~\bibnamefont
  {Vanderhaeghen}},\ }\href {\doibase 10.1016/j.nuclphysa.2004.06.008}
  {\bibfield  {journal} {\bibinfo  {journal} {Nucl. Phys.}\ }\textbf {\bibinfo
  {volume} {A741}},\ \bibinfo {pages} {234} (\bibinfo {year}
  {2004})}\BibitemShut {NoStop}%
\bibitem [{\citenamefont {Gorchtein}(2006{\natexlab{a}})}]{Gorchtein:2005yz}%
  \BibitemOpen
  \bibfield  {author} {\bibinfo {author} {\bibfnamefont {M.}~\bibnamefont
  {Gorchtein}},\ }\href {\doibase 10.1103/PhysRevC.73.055201} {\bibfield
  {journal} {\bibinfo  {journal} {Phys. Rev.}\ }\textbf {\bibinfo {volume}
  {C73}},\ \bibinfo {pages} {055201} (\bibinfo {year}
  {2006}{\natexlab{a}})}\BibitemShut {NoStop}%
\bibitem [{\citenamefont {Gorchtein}(2006{\natexlab{b}})}]{PhysRevC.73.035213}%
  \BibitemOpen
  \bibfield  {author} {\bibinfo {author} {\bibfnamefont {M.}~\bibnamefont
  {Gorchtein}},\ }\href {\doibase 10.1103/PhysRevC.73.035213} {\bibfield
  {journal} {\bibinfo  {journal} {Phys. Rev. C}\ }\textbf {\bibinfo {volume}
  {73}},\ \bibinfo {pages} {035213} (\bibinfo {year}
  {2006}{\natexlab{b}})}\BibitemShut {NoStop}%
\bibitem [{\citenamefont {Diaconescu}\ and\ \citenamefont
  {Ramsey-Musolf}(2004)}]{Diaconescu:2004aa}%
  \BibitemOpen
  \bibfield  {author} {\bibinfo {author} {\bibfnamefont {L.}~\bibnamefont
  {Diaconescu}}\ and\ \bibinfo {author} {\bibfnamefont {M.}~\bibnamefont
  {Ramsey-Musolf}},\ }\href {\doibase 10.1103/PhysRevC.70.054003} {\bibfield
  {journal} {\bibinfo  {journal} {Phys. Rev. C}\ }\textbf {\bibinfo {volume}
  {70}},\ \bibinfo {pages} {054003} (\bibinfo {year} {2004})}\BibitemShut
  {NoStop}%
\bibitem [{\citenamefont {Armstrong}\ \emph {et~al.}(2007)\citenamefont
  {Armstrong} \emph {et~al.}}]{Armstrong:2007vm}%
  \BibitemOpen
  \bibfield  {author} {\bibinfo {author} {\bibfnamefont {D.~S.}\ \bibnamefont
  {Armstrong}} \emph {et~al.} (\bibinfo {collaboration} {G0 Collaboration}),\
  }\href {\doibase 10.1103/PhysRevLett.99.092301} {\bibfield  {journal}
  {\bibinfo  {journal} {Phys. Rev. Lett.}\ }\textbf {\bibinfo {volume} {99}},\
  \bibinfo {pages} {092301} (\bibinfo {year} {2007})}\BibitemShut {NoStop}%
\bibitem [{\citenamefont {Androi\'c}\ \emph {et~al.}(2011)\citenamefont
  {Androi\'c} \emph {et~al.}}]{Androic:2011rh}%
  \BibitemOpen
  \bibfield  {author} {\bibinfo {author} {\bibfnamefont {D.}~\bibnamefont
  {Androi\'c}} \emph {et~al.} (\bibinfo {collaboration} {G0 Collaboration}),\
  }\href {\doibase 10.1103/PhysRevLett.107.022501} {\bibfield  {journal}
  {\bibinfo  {journal} {Phys. Rev. Lett.}\ }\textbf {\bibinfo {volume} {107}},\
  \bibinfo {pages} {022501} (\bibinfo {year} {2011})}\BibitemShut {NoStop}%
\bibitem [{\citenamefont {Maas}\ \emph {et~al.}(2005)\citenamefont {Maas} \emph
  {et~al.}}]{Maas:2004pd}%
  \BibitemOpen
  \bibfield  {author} {\bibinfo {author} {\bibfnamefont {F.~E.}\ \bibnamefont
  {Maas}} \emph {et~al.},\ }\href {\doibase 10.1103/PhysRevLett.94.082001}
  {\bibfield  {journal} {\bibinfo  {journal} {Phys. Rev. Lett.}\ }\textbf
  {\bibinfo {volume} {94}},\ \bibinfo {pages} {082001} (\bibinfo {year}
  {2005})}\BibitemShut {NoStop}%
\bibitem [{\citenamefont {Gou}\ \emph {et~al.}(2020)\citenamefont {Gou} \emph
  {et~al.}}]{Gou:2020viq}%
  \BibitemOpen
  \bibfield  {author} {\bibinfo {author} {\bibfnamefont {B.}~\bibnamefont
  {Gou}} \emph {et~al.},\ }\href {\doibase 10.1103/PhysRevLett.124.122003}
  {\bibfield  {journal} {\bibinfo  {journal} {Phys.\ Rev.\ Lett.}\ }\textbf
  {\bibinfo {volume} {124}},\ \bibinfo {pages} {122003} (\bibinfo {year}
  {2020})}\BibitemShut {NoStop}%
\bibitem [{\citenamefont {R\'{\i}os}\ \emph {et~al.}(2017)\citenamefont
  {R\'{\i}os} \emph {et~al.}}]{Rios:2017vsw}%
  \BibitemOpen
  \bibfield  {author} {\bibinfo {author} {\bibfnamefont {D.~B.}\ \bibnamefont
  {R\'{\i}os}} \emph {et~al.},\ }\href {\doibase
  10.1103/PhysRevLett.119.012501} {\bibfield  {journal} {\bibinfo  {journal}
  {Phys. Rev. Lett.}\ }\textbf {\bibinfo {volume} {119}},\ \bibinfo {pages}
  {012501} (\bibinfo {year} {2017})}\BibitemShut {NoStop}%
\bibitem [{\citenamefont {Androi\'c}\ \emph {et~al.}(2020)\citenamefont
  {Androi\'c} \emph {et~al.}}]{Androic:2020rkw}%
  \BibitemOpen
  \bibfield  {author} {\bibinfo {author} {\bibfnamefont {D.}~\bibnamefont
  {Androi\'c}} \emph {et~al.} (\bibinfo {collaboration} {Q$_{\rm weak}$
  Collaboration}),\ }\href {\doibase 10.1103/PhysRevLett.125.112502} {\bibfield
   {journal} {\bibinfo  {journal} {Phys. Rev. Lett.}\ }\textbf {\bibinfo
  {volume} {125}},\ \bibinfo {pages} {112502} (\bibinfo {year}
  {2020})}\BibitemShut {NoStop}%
\bibitem [{\citenamefont {Souder}\ and\ \citenamefont
  {Paschke}(2016)}]{Souder:2015mlu}%
  \BibitemOpen
  \bibfield  {author} {\bibinfo {author} {\bibfnamefont {P.}~\bibnamefont
  {Souder}}\ and\ \bibinfo {author} {\bibfnamefont {K.~D.}\ \bibnamefont
  {Paschke}},\ }\href {\doibase 10.1007/s11467-015-0482-0} {\bibfield
  {journal} {\bibinfo  {journal} {Front. Phys. (Beijing)}\ }\textbf {\bibinfo
  {volume} {11}},\ \bibinfo {pages} {111301} (\bibinfo {year}
  {2016})}\BibitemShut {NoStop}%
\bibitem [{\citenamefont {Armstrong}\ and\ \citenamefont
  {McKeown}(2012)}]{Armstrong:2012bi}%
  \BibitemOpen
  \bibfield  {author} {\bibinfo {author} {\bibfnamefont {D.~S.}\ \bibnamefont
  {Armstrong}}\ and\ \bibinfo {author} {\bibfnamefont {R.~D.}\ \bibnamefont
  {McKeown}},\ }\href {\doibase 10.1146/annurev-nucl-102010-130419} {\bibfield
  {journal} {\bibinfo  {journal} {Ann. Rev. Nucl. Part. Sci.}\ }\textbf
  {\bibinfo {volume} {62}},\ \bibinfo {pages} {337} (\bibinfo {year}
  {2012})}\BibitemShut {NoStop}%
\bibitem [{\citenamefont {Carlini}\ \emph {et~al.}(2019)\citenamefont
  {Carlini}, \citenamefont {van Oers}, \citenamefont {Pitt},\ and\
  \citenamefont {Smith}}]{Carlini2019}%
  \BibitemOpen
  \bibfield  {author} {\bibinfo {author} {\bibfnamefont {R.~D.}\ \bibnamefont
  {Carlini}}, \bibinfo {author} {\bibfnamefont {W.~T.}\ \bibnamefont {van
  Oers}}, \bibinfo {author} {\bibfnamefont {M.~L.}\ \bibnamefont {Pitt}}, \
  and\ \bibinfo {author} {\bibfnamefont {G.~R.}\ \bibnamefont {Smith}},\ }\href
  {\doibase 10.1146/annurev-nucl-101918-023633} {\bibfield  {journal} {\bibinfo
   {journal} {Ann. Rev. Nucl. Part. Sci.}\ }\textbf {\bibinfo {volume} {69}},\
  \bibinfo {pages} {191} (\bibinfo {year} {2019})}\BibitemShut {NoStop}%
\bibitem [{\citenamefont {Zhang}\ \emph {et~al.}(2015)\citenamefont {Zhang}
  \emph {et~al.}}]{PhysRevLett.115.172502}%
  \BibitemOpen
  \bibfield  {author} {\bibinfo {author} {\bibfnamefont {Y.~W.}\ \bibnamefont
  {Zhang}} \emph {et~al.} (\bibinfo {collaboration} {Jefferson Lab Hall A
  Collaboration}),\ }\href {\doibase 10.1103/PhysRevLett.115.172502} {\bibfield
   {journal} {\bibinfo  {journal} {Phys. Rev. Lett.}\ }\textbf {\bibinfo
  {volume} {115}},\ \bibinfo {pages} {172502} (\bibinfo {year}
  {2015})}\BibitemShut {NoStop}%
\bibitem [{\citenamefont {Chen}\ \emph {et~al.}(2004)\citenamefont {Chen},
  \citenamefont {Afanasev}, \citenamefont {Brodsky}, \citenamefont {Carlson},\
  and\ \citenamefont {Vanderhaeghen}}]{PhysRevLett.93.122301}%
  \BibitemOpen
  \bibfield  {author} {\bibinfo {author} {\bibfnamefont {Y.-C.}\ \bibnamefont
  {Chen}}, \bibinfo {author} {\bibfnamefont {A.}~\bibnamefont {Afanasev}},
  \bibinfo {author} {\bibfnamefont {S.~J.}\ \bibnamefont {Brodsky}}, \bibinfo
  {author} {\bibfnamefont {C.~E.}\ \bibnamefont {Carlson}}, \ and\ \bibinfo
  {author} {\bibfnamefont {M.}~\bibnamefont {Vanderhaeghen}},\ }\href {\doibase
  10.1103/PhysRevLett.93.122301} {\bibfield  {journal} {\bibinfo  {journal}
  {Phys. Rev. Lett.}\ }\textbf {\bibinfo {volume} {93}},\ \bibinfo {pages}
  {122301} (\bibinfo {year} {2004})}\BibitemShut {NoStop}%
\bibitem [{\citenamefont {Abrahamyan}\ \emph {et~al.}(2012)\citenamefont
  {Abrahamyan} \emph {et~al.}}]{Abrahamyan:2012cg}%
  \BibitemOpen
  \bibfield  {author} {\bibinfo {author} {\bibfnamefont {S.}~\bibnamefont
  {Abrahamyan}} \emph {et~al.} (\bibinfo {collaboration} {HAPPEX, PREX
  Collaborations}),\ }\href {\doibase 10.1103/PhysRevLett.109.192501}
  {\bibfield  {journal} {\bibinfo  {journal} {Phys. Rev. Lett.}\ }\textbf
  {\bibinfo {volume} {109}},\ \bibinfo {pages} {192501} (\bibinfo {year}
  {2012})}\BibitemShut {NoStop}%
\bibitem [{\citenamefont {Esser}\ \emph {et~al.}(2018)\citenamefont {Esser}
  \emph {et~al.}}]{Esser:2018vdp}%
  \BibitemOpen
  \bibfield  {author} {\bibinfo {author} {\bibfnamefont {A.}~\bibnamefont
  {Esser}} \emph {et~al.},\ }\href {\doibase 10.1103/PhysRevLett.121.022503}
  {\bibfield  {journal} {\bibinfo  {journal} {Phys. Rev. Lett.}\ }\textbf
  {\bibinfo {volume} {121}},\ \bibinfo {pages} {022503} (\bibinfo {year}
  {2018})}\BibitemShut {NoStop}%
\bibitem [{\citenamefont {Esser}\ \emph {et~al.}(2020)\citenamefont {Esser}
  \emph {et~al.}}]{Esser:2020vjb}%
  \BibitemOpen
  \bibfield  {author} {\bibinfo {author} {\bibfnamefont {A.}~\bibnamefont
  {Esser}} \emph {et~al.},\ }\href {\doibase
  https://doi.org/10.1016/j.physletb.2020.135664} {\bibfield  {journal}
  {\bibinfo  {journal} {Phys. Lett. B}\ }\textbf {\bibinfo {volume} {808}},\
  \bibinfo {pages} {135664} (\bibinfo {year} {2020})}\BibitemShut {NoStop}%
\bibitem [{\citenamefont {Cooper}\ and\ \citenamefont
  {Horowitz}(2005)}]{Cooper:2005sk}%
  \BibitemOpen
  \bibfield  {author} {\bibinfo {author} {\bibfnamefont {E.~D.}\ \bibnamefont
  {Cooper}}\ and\ \bibinfo {author} {\bibfnamefont {C.~J.}\ \bibnamefont
  {Horowitz}},\ }\href {\doibase 10.1103/PhysRevC.72.034602} {\bibfield
  {journal} {\bibinfo  {journal} {Phys. Rev. C}\ }\textbf {\bibinfo {volume}
  {72}},\ \bibinfo {pages} {034602} (\bibinfo {year} {2005})}\BibitemShut
  {NoStop}%
\bibitem [{\citenamefont {Gorchtein}\ and\ \citenamefont
  {Horowitz}(2008)}]{Gorchtein:2008dy}%
  \BibitemOpen
  \bibfield  {author} {\bibinfo {author} {\bibfnamefont {M.}~\bibnamefont
  {Gorchtein}}\ and\ \bibinfo {author} {\bibfnamefont {C.~J.}\ \bibnamefont
  {Horowitz}},\ }\href {\doibase 10.1103/PhysRevC.77.044606} {\bibfield
  {journal} {\bibinfo  {journal} {Phys. Rev. C}\ }\textbf {\bibinfo {volume}
  {77}},\ \bibinfo {pages} {044606} (\bibinfo {year} {2008})}\BibitemShut
  {NoStop}%
\bibitem [{\citenamefont {Gorchtein}(2006{\natexlab{c}})}]{Gorchtein:2005za}%
  \BibitemOpen
  \bibfield  {author} {\bibinfo {author} {\bibfnamefont {M.}~\bibnamefont
  {Gorchtein}},\ }\href {\doibase 10.1103/PhysRevC.73.035213} {\bibfield
  {journal} {\bibinfo  {journal} {Phys. Rev. C}\ }\textbf {\bibinfo {volume}
  {73}},\ \bibinfo {pages} {035213} (\bibinfo {year}
  {2006}{\natexlab{c}})}\BibitemShut {NoStop}%
\bibitem [{\citenamefont {Bianchi}\ \emph {et~al.}(1996)\citenamefont {Bianchi}
  \emph {et~al.}}]{Bianchi:1995vb}%
  \BibitemOpen
  \bibfield  {author} {\bibinfo {author} {\bibfnamefont {N.}~\bibnamefont
  {Bianchi}} \emph {et~al.},\ }\href {\doibase 10.1103/PhysRevC.54.1688}
  {\bibfield  {journal} {\bibinfo  {journal} {Phys. Rev. C}\ }\textbf {\bibinfo
  {volume} {54}},\ \bibinfo {pages} {1688} (\bibinfo {year}
  {1996})}\BibitemShut {NoStop}%
\bibitem [{\citenamefont {Bauer}\ \emph {et~al.}(1978)\citenamefont {Bauer},
  \citenamefont {Spital}, \citenamefont {Yennie},\ and\ \citenamefont
  {Pipkin}}]{Bauer:1977iq}%
  \BibitemOpen
  \bibfield  {author} {\bibinfo {author} {\bibfnamefont {T.~H.}\ \bibnamefont
  {Bauer}}, \bibinfo {author} {\bibfnamefont {R.~D.}\ \bibnamefont {Spital}},
  \bibinfo {author} {\bibfnamefont {D.~R.}\ \bibnamefont {Yennie}}, \ and\
  \bibinfo {author} {\bibfnamefont {F.~M.}\ \bibnamefont {Pipkin}},\ }\href
  {\doibase 10.1103/RevModPhys.50.261} {\bibfield  {journal} {\bibinfo
  {journal} {Rev. Mod. Phys.}\ }\textbf {\bibinfo {volume} {50}},\ \bibinfo
  {pages} {261} (\bibinfo {year} {1978})},\ \bibinfo {note} {[Erratum:
  Rev.Mod.Phys. 51, 407 (1979)]}\BibitemShut {NoStop}%
\bibitem [{\citenamefont {Aleksanian}\ \emph {et~al.}(1987)\citenamefont
  {Aleksanian} \emph {et~al.}}]{Aleksanian:1986hb}%
  \BibitemOpen
  \bibfield  {author} {\bibinfo {author} {\bibfnamefont {A.~S.}\ \bibnamefont
  {Aleksanian}} \emph {et~al.},\ }\href@noop {} {\bibfield  {journal} {\bibinfo
   {journal} {Sov. J. Nucl. Phys.}\ }\textbf {\bibinfo {volume} {45}},\
  \bibinfo {pages} {628} (\bibinfo {year} {1987})}\BibitemShut {NoStop}%
\bibitem [{\citenamefont {McNulty}(2020)}]{McNulty:2020}%
  \BibitemOpen
  \bibfield  {author} {\bibinfo {author} {\bibfnamefont {D.}~\bibnamefont
  {McNulty}},\ }in\ \href@noop {} {\emph {\bibinfo {booktitle} {Parity
  Violation and related topics, MITP virtual workshop}}}\ (\bibinfo {year}
  {2020})\BibitemShut {NoStop}%
\bibitem [{\citenamefont {Richards}(2020)}]{Richards:2020}%
  \BibitemOpen
  \bibfield  {author} {\bibinfo {author} {\bibfnamefont {R.}~\bibnamefont
  {Richards}},\ }in\ \href@noop {} {\emph {\bibinfo {booktitle} {APS/DNP 2020,
  Bull. Am. Phys. Soc. FM.00006}}}\ (\bibinfo {year} {2020})\BibitemShut
  {NoStop}%
\bibitem [{\citenamefont {Allison}\ \emph {et~al.}(2015)\citenamefont {Allison}
  \emph {et~al.}}]{ALLISON2015105}%
  \BibitemOpen
  \bibfield  {author} {\bibinfo {author} {\bibfnamefont {T.}~\bibnamefont
  {Allison}} \emph {et~al.} (\bibinfo {collaboration} {Q$_{\rm weak}$
  Collaboration}),\ }\href {\doibase
  https://doi.org/10.1016/j.nima.2015.01.023} {\bibfield  {journal} {\bibinfo
  {journal} {Nucl. Instrum. Methods}\ }\textbf {\bibinfo {volume} {A781}},\
  \bibinfo {pages} {105 } (\bibinfo {year} {2015})}\BibitemShut {NoStop}%
\bibitem [{\citenamefont {Adderley}\ \emph {et~al.}(2011)\citenamefont
  {Adderley}, \citenamefont {Benesch}, \citenamefont {Clark}, \citenamefont
  {Grames}, \citenamefont {Hansknecht} \emph {et~al.}}]{Adderley:2011ri}%
  \BibitemOpen
  \bibfield  {author} {\bibinfo {author} {\bibfnamefont {P.~A.}\ \bibnamefont
  {Adderley}}, \bibinfo {author} {\bibfnamefont {J.~F.}\ \bibnamefont
  {Benesch}}, \bibinfo {author} {\bibfnamefont {J.}~\bibnamefont {Clark}},
  \bibinfo {author} {\bibfnamefont {J.~M.}\ \bibnamefont {Grames}}, \bibinfo
  {author} {\bibfnamefont {J.}~\bibnamefont {Hansknecht}},  \emph {et~al.},\
  }\href@noop {} {\bibfield  {journal} {\bibinfo  {journal} {Conf.Proc.}\
  }\textbf {\bibinfo {volume} {C110328}},\ \bibinfo {pages} {862} (\bibinfo
  {year} {2011})}\BibitemShut {NoStop}%
\bibitem [{\citenamefont {Yan}\ \emph {et~al.}(1995)\citenamefont {Yan} \emph
  {et~al.}}]{Yan:1995nc}%
  \BibitemOpen
  \bibfield  {author} {\bibinfo {author} {\bibfnamefont {C.}~\bibnamefont
  {Yan}} \emph {et~al.},\ }\href {\doibase 10.1016/0168-9002(95)00505-6}
  {\bibfield  {journal} {\bibinfo  {journal} {Nucl. Instrum. Meth. A}\ }\textbf
  {\bibinfo {volume} {365}},\ \bibinfo {pages} {261} (\bibinfo {year}
  {1995})}\BibitemShut {NoStop}%
\bibitem [{\citenamefont {Magee}\ \emph {et~al.}(2017)\citenamefont {Magee}
  \emph {et~al.}}]{Magee:2016xqx}%
  \BibitemOpen
  \bibfield  {author} {\bibinfo {author} {\bibfnamefont {J.~A.}\ \bibnamefont
  {Magee}} \emph {et~al.},\ }\href {\doibase 10.1016/j.physletb.2017.01.026}
  {\bibfield  {journal} {\bibinfo  {journal} {Phys. Lett.}\ }\textbf {\bibinfo
  {volume} {B766}},\ \bibinfo {pages} {339} (\bibinfo {year}
  {2017})}\BibitemShut {NoStop}%
\bibitem [{\citenamefont {Narayan}\ \emph {et~al.}(2016)\citenamefont {Narayan}
  \emph {et~al.}}]{Narayan:2015aua}%
  \BibitemOpen
  \bibfield  {author} {\bibinfo {author} {\bibfnamefont {A.}~\bibnamefont
  {Narayan}} \emph {et~al.},\ }\href {\doibase 10.1103/PhysRevX.6.011013}
  {\bibfield  {journal} {\bibinfo  {journal} {Phys. Rev.}\ }\textbf {\bibinfo
  {volume} {X6}},\ \bibinfo {pages} {011013} (\bibinfo {year}
  {2016})}\BibitemShut {NoStop}%
\bibitem [{\citenamefont {Hauger}\ \emph {et~al.}(2001)\citenamefont {Hauger}
  \emph {et~al.}}]{Hauger:1999iv}%
  \BibitemOpen
  \bibfield  {author} {\bibinfo {author} {\bibfnamefont {M.}~\bibnamefont
  {Hauger}} \emph {et~al.},\ }\href {\doibase 10.1016/S0168-9002(01)00197-8}
  {\bibfield  {journal} {\bibinfo  {journal} {Nucl. Instrum. Methods}\ }\textbf
  {\bibinfo {volume} {A462}},\ \bibinfo {pages} {382} (\bibinfo {year}
  {2001})}\BibitemShut {NoStop}%
\bibitem [{ATS()}]{ATS}%
  \BibitemOpen
  \href@noop {} {}\bibinfo {note} {Applied Technical Services, Atlanta,
  GA}\BibitemShut {NoStop}%
\bibitem [{\citenamefont {Agostinelli}\ \emph {et~al.}(2003)\citenamefont
  {Agostinelli} \emph {et~al.}}]{Agostinelli:2002hh}%
  \BibitemOpen
  \bibfield  {author} {\bibinfo {author} {\bibfnamefont {S.}~\bibnamefont
  {Agostinelli}} \emph {et~al.} (\bibinfo {collaboration} {GEANT4}),\ }\href
  {\doibase 10.1016/S0168-9002(03)01368-8} {\bibfield  {journal} {\bibinfo
  {journal} {Nucl. Instrum. Meth. A}\ }\textbf {\bibinfo {volume} {506}},\
  \bibinfo {pages} {250} (\bibinfo {year} {2003})}\BibitemShut {NoStop}%
\bibitem [{\citenamefont {Pan}\ \emph {et~al.}(2016)\citenamefont {Pan} \emph
  {et~al.}}]{Pan:2016rbx}%
  \BibitemOpen
  \bibfield  {author} {\bibinfo {author} {\bibfnamefont {J.}~\bibnamefont
  {Pan}} \emph {et~al.},\ }\href {\doibase 10.1007/s10751-016-1369-3}
  {\bibfield  {journal} {\bibinfo  {journal} {Hyperfine Interact.}\ }\textbf
  {\bibinfo {volume} {237}},\ \bibinfo {pages} {161} (\bibinfo {year}
  {2016})}\BibitemShut {NoStop}%
\bibitem [{\citenamefont {McHugh}(2017)}]{marty-thesis}%
  \BibitemOpen
  \bibfield  {author} {\bibinfo {author} {\bibfnamefont {M.~J.}\ \bibnamefont
  {McHugh}},\ }\emph {\bibinfo {title} {A Measurement of the Transverse
  Asymmetry in Forward-Angle Electron-Carbon Scattering Using the Qweak
  Apparatus}},\ \href
  {https://misportal.jlab.org/ul/publications/view_pub.cfm?pub_id=14918}
  {\bibinfo {type} {{PhD} dissertation}},\ \bibinfo  {school} {George
  Washington U.} (\bibinfo {year} {2017})\BibitemShut {NoStop}%
\bibitem [{\citenamefont {Bartlett}(2018)}]{kurtis-thesis}%
  \BibitemOpen
  \bibfield  {author} {\bibinfo {author} {\bibfnamefont {K.~D.}\ \bibnamefont
  {Bartlett}},\ }\emph {\bibinfo {title} {First Measurements of the
  Parity-Violating and Beam-Normal Single-Spin Asymmetries in Elastic
  Electron-Aluminum Scattering}},\ \href
  {https://misportal.jlab.org/ul/publications/view_pub.cfm?pub_id=15621}
  {\bibinfo {type} {{PhD} dissertation}},\ \bibinfo  {school} {William \& Mary}
  (\bibinfo {year} {2018})\BibitemShut {NoStop}%
\bibitem [{\citenamefont {Waidyawansa}(2013)}]{BWaidyawansa_phd}%
  \BibitemOpen
  \bibfield  {author} {\bibinfo {author} {\bibfnamefont {D.~B.~P.}\
  \bibnamefont {Waidyawansa}},\ }\emph {\bibinfo {title} {A 3\% Measurement of
  the Beam Normal Single Spin Asymmetry in Forward Angle Elastic Electron
  Proton Scattering Using the Q$_{\rm weak}$ Setup}},\ \href
  {https://misportal.jlab.org/ul/publications/downloadFile.cfm?pub_id=12540}
  {Ph.D. thesis},\ \bibinfo  {school} {Ohio University} (\bibinfo {year}
  {2013})\BibitemShut {NoStop}%
\bibitem [{\citenamefont {Duvall}(2017)}]{Duvall_phd}%
  \BibitemOpen
  \bibfield  {author} {\bibinfo {author} {\bibfnamefont {W.}~\bibnamefont
  {Duvall}},\ }\emph {\bibinfo {title} {Precision Measurement of the Proton's
  Weak Charge using Parity-Violating Electron Scattering}},\ \href
  {https://misportal.jlab.org/ul/publications/downloadFile.cfm?pub_id=15805}
  {Ph.D. thesis},\ \bibinfo  {school} {Virginia Tech} (\bibinfo {year}
  {2017})\BibitemShut {NoStop}%
\bibitem [{\citenamefont {Horowitz}\ and\ \citenamefont {Lin}(2017)}]{chuck}%
  \BibitemOpen
  \bibfield  {author} {\bibinfo {author} {\bibfnamefont {C.~J.}\ \bibnamefont
  {Horowitz}}\ and\ \bibinfo {author} {\bibfnamefont {Z.}~\bibnamefont {Lin}},\
  }\href@noop {} {}\bibinfo {howpublished} {private communication} (\bibinfo
  {year} {2017})\BibitemShut {NoStop}%
\bibitem [{\citenamefont {De~Vries}\ \emph {et~al.}(1987)\citenamefont
  {De~Vries}, \citenamefont {De~Jager},\ and\ \citenamefont
  {De~Vries}}]{DEVRIES1987495}%
  \BibitemOpen
  \bibfield  {author} {\bibinfo {author} {\bibfnamefont {H.}~\bibnamefont
  {De~Vries}}, \bibinfo {author} {\bibfnamefont {C.}~\bibnamefont {De~Jager}},
  \ and\ \bibinfo {author} {\bibfnamefont {C.}~\bibnamefont {De~Vries}},\
  }\href {\doibase https://doi.org/10.1016/0092-640X(87)90013-1} {\bibfield
  {journal} {\bibinfo  {journal} {Atomic Data and Nuclear Data Tables}\
  }\textbf {\bibinfo {volume} {36}},\ \bibinfo {pages} {495 } (\bibinfo {year}
  {1987})}\BibitemShut {NoStop}%
\bibitem [{\citenamefont {Wiser}(1977)}]{Wiser}%
  \BibitemOpen
  \bibfield  {author} {\bibinfo {author} {\bibfnamefont {D.~E.}\ \bibnamefont
  {Wiser}},\ }\emph {\bibinfo {title} {Inclusive Photoproduction of Protons,
  Kaons, and Pions at SLAC Energies}},\ \href@noop {} {\bibinfo {type} {{PhD}
  dissertation}},\ \bibinfo  {school} {University of Wisconsin-Madison}
  (\bibinfo {year} {1977})\BibitemShut {NoStop}%
\bibitem [{\citenamefont {Christy}()}]{Christy_future}%
  \BibitemOpen
  \bibfield  {author} {\bibinfo {author} {\bibfnamefont {M.~E.}\ \bibnamefont
  {Christy}},\ }\href@noop {} {}\bibinfo {howpublished} {to be
  published}\BibitemShut {NoStop}%
\bibitem [{\citenamefont {Donnelly}\ and\ \citenamefont
  {Sick}(1999)}]{donnelly-sick:99}%
  \BibitemOpen
  \bibfield  {author} {\bibinfo {author} {\bibfnamefont {T.~W.}\ \bibnamefont
  {Donnelly}}\ and\ \bibinfo {author} {\bibfnamefont {I.}~\bibnamefont
  {Sick}},\ }\href {\doibase 10.1103/PhysRevC.60.065502} {\bibfield  {journal}
  {\bibinfo  {journal} {Phys. Rev. C}\ }\textbf {\bibinfo {volume} {60}},\
  \bibinfo {pages} {065502} (\bibinfo {year} {1999})}\BibitemShut {NoStop}%
\bibitem [{\citenamefont {Bosted}\ and\ \citenamefont
  {Mamyan}(2012)}]{Bosted:2012qc}%
  \BibitemOpen
  \bibfield  {author} {\bibinfo {author} {\bibfnamefont {P.~E.}\ \bibnamefont
  {Bosted}}\ and\ \bibinfo {author} {\bibfnamefont {V.}~\bibnamefont
  {Mamyan}},\ }\href@noop {} {\  (\bibinfo {year} {2012})},\ \Eprint
  {http://arxiv.org/abs/1203.2262} {arXiv:1203.2262 [nucl-th]} \BibitemShut
  {NoStop}%
\bibitem [{\citenamefont {Nuruzzaman}(2014)}]{Nuruzzaman_phd}%
  \BibitemOpen
  \bibfield  {author} {\bibinfo {author} {\bibnamefont {Nuruzzaman}},\ }\emph
  {\bibinfo {title} {Beam Normal Single Spin Asymmetry in Forward Angle
  Inelastic Electron-Proton Scattering using the Q-Weak Apparatus}},\ \href
  {https://misportal.jlab.org/ul/publications/downloadFile.cfm?pub_id=13632}
  {Ph.D. thesis},\ \bibinfo  {school} {Hampton University} (\bibinfo {year}
  {2014})\BibitemShut {NoStop}%
\bibitem [{\citenamefont {Nuruzzaman}(2015)}]{Nuruzzaman:2015vba}%
  \BibitemOpen
  \bibfield  {author} {\bibinfo {author} {\bibnamefont {Nuruzzaman}} (\bibinfo
  {collaboration} {Q$_{\rm weak}$ Collaboration}),\ }in\ \href@noop {} {\emph
  {\bibinfo {booktitle} {{Proceedings, CIPANP 2015}}}}\ (\bibinfo {year}
  {2015})\ \Eprint {http://arxiv.org/abs/1510.00449} {arXiv:1510.00449
  [nucl-ex]} \BibitemShut {NoStop}%
\bibitem [{\citenamefont {Carlson}\ \emph {et~al.}(2017)\citenamefont
  {Carlson}, \citenamefont {Pasquini}, \citenamefont {Pauk},\ and\
  \citenamefont {Vanderhaeghen}}]{Carlson:2017lys}%
  \BibitemOpen
  \bibfield  {author} {\bibinfo {author} {\bibfnamefont {C.~E.}\ \bibnamefont
  {Carlson}}, \bibinfo {author} {\bibfnamefont {B.}~\bibnamefont {Pasquini}},
  \bibinfo {author} {\bibfnamefont {V.}~\bibnamefont {Pauk}}, \ and\ \bibinfo
  {author} {\bibfnamefont {M.}~\bibnamefont {Vanderhaeghen}},\ }\href {\doibase
  10.1103/PhysRevD.96.113010} {\bibfield  {journal} {\bibinfo  {journal} {Phys.
  Rev.}\ }\textbf {\bibinfo {volume} {D96}},\ \bibinfo {pages} {113010}
  (\bibinfo {year} {2017})}\BibitemShut {NoStop}%
\bibitem [{\citenamefont {Singhal}\ \emph {et~al.}(1977)\citenamefont
  {Singhal}, \citenamefont {Johnston}, \citenamefont {Gillespie},\ and\
  \citenamefont {Lees}}]{Singhal:1977wvn}%
  \BibitemOpen
  \bibfield  {author} {\bibinfo {author} {\bibfnamefont {R.~P.}\ \bibnamefont
  {Singhal}}, \bibinfo {author} {\bibfnamefont {A.}~\bibnamefont {Johnston}},
  \bibinfo {author} {\bibfnamefont {W.~A.}\ \bibnamefont {Gillespie}}, \ and\
  \bibinfo {author} {\bibfnamefont {E.~W.}\ \bibnamefont {Lees}},\ }\href
  {\doibase 10.1016/0375-9474(77)90418-3} {\bibfield  {journal} {\bibinfo
  {journal} {Nucl. Phys. A}\ }\textbf {\bibinfo {volume} {279}},\ \bibinfo
  {pages} {29} (\bibinfo {year} {1977})}\BibitemShut {NoStop}%
\bibitem [{\citenamefont {Hicks}\ \emph {et~al.}(1980)\citenamefont {Hicks},
  \citenamefont {Hotta}, \citenamefont {Flanz},\ and\ \citenamefont
  {De~Vries}}]{Hicks:1980bv}%
  \BibitemOpen
  \bibfield  {author} {\bibinfo {author} {\bibfnamefont {R.~S.}\ \bibnamefont
  {Hicks}}, \bibinfo {author} {\bibfnamefont {A.}~\bibnamefont {Hotta}},
  \bibinfo {author} {\bibfnamefont {J.~B.}\ \bibnamefont {Flanz}}, \ and\
  \bibinfo {author} {\bibfnamefont {H.}~\bibnamefont {De~Vries}},\ }\href
  {\doibase 10.1103/PhysRevC.21.2177} {\bibfield  {journal} {\bibinfo
  {journal} {Phys. Rev. C}\ }\textbf {\bibinfo {volume} {21}},\ \bibinfo
  {pages} {2177} (\bibinfo {year} {1980})}\BibitemShut {NoStop}%
\bibitem [{\citenamefont {Ryan}\ \emph {et~al.}(1983)\citenamefont {Ryan},
  \citenamefont {Hicks}, \citenamefont {Hotta}, \citenamefont {Dubach},
  \citenamefont {Peterson},\ and\ \citenamefont {Webb}}]{Ryan:1983zz}%
  \BibitemOpen
  \bibfield  {author} {\bibinfo {author} {\bibfnamefont {P.~J.}\ \bibnamefont
  {Ryan}}, \bibinfo {author} {\bibfnamefont {R.~S.}\ \bibnamefont {Hicks}},
  \bibinfo {author} {\bibfnamefont {A.}~\bibnamefont {Hotta}}, \bibinfo
  {author} {\bibfnamefont {J.~F.}\ \bibnamefont {Dubach}}, \bibinfo {author}
  {\bibfnamefont {G.~A.}\ \bibnamefont {Peterson}}, \ and\ \bibinfo {author}
  {\bibfnamefont {D.~V.}\ \bibnamefont {Webb}},\ }\href {\doibase
  10.1103/PhysRevC.27.2515} {\bibfield  {journal} {\bibinfo  {journal} {Phys.
  Rev. C}\ }\textbf {\bibinfo {volume} {27}},\ \bibinfo {pages} {2515}
  (\bibinfo {year} {1983})}\BibitemShut {NoStop}%
\bibitem [{\citenamefont {Crannell}\ and\ \citenamefont
  {Griffy}(1964)}]{Crannell:1964zz}%
  \BibitemOpen
  \bibfield  {author} {\bibinfo {author} {\bibfnamefont {H.~L.}\ \bibnamefont
  {Crannell}}\ and\ \bibinfo {author} {\bibfnamefont {T.~A.}\ \bibnamefont
  {Griffy}},\ }\href {\doibase 10.1103/PhysRev.136.B1580} {\bibfield  {journal}
  {\bibinfo  {journal} {Phys. Rev.}\ }\textbf {\bibinfo {volume} {136}},\
  \bibinfo {pages} {B1580} (\bibinfo {year} {1964})}\BibitemShut {NoStop}%
\bibitem [{\citenamefont {Crannell}(1966)}]{PhysRev.148.1107}%
  \BibitemOpen
  \bibfield  {author} {\bibinfo {author} {\bibfnamefont {H.}~\bibnamefont
  {Crannell}},\ }\href {\doibase 10.1103/PhysRev.148.1107} {\bibfield
  {journal} {\bibinfo  {journal} {Phys. Rev.}\ }\textbf {\bibinfo {volume}
  {148}},\ \bibinfo {pages} {1107} (\bibinfo {year} {1966})}\BibitemShut
  {NoStop}%
\bibitem [{\citenamefont {Nakada}\ \emph {et~al.}(1971)\citenamefont {Nakada},
  \citenamefont {Torizuka},\ and\ \citenamefont
  {Horikawa}}]{PhysRevLett.27.745}%
  \BibitemOpen
  \bibfield  {author} {\bibinfo {author} {\bibfnamefont {A.}~\bibnamefont
  {Nakada}}, \bibinfo {author} {\bibfnamefont {Y.}~\bibnamefont {Torizuka}}, \
  and\ \bibinfo {author} {\bibfnamefont {Y.}~\bibnamefont {Horikawa}},\ }\href
  {\doibase 10.1103/PhysRevLett.27.745} {\bibfield  {journal} {\bibinfo
  {journal} {Phys. Rev. Lett.}\ }\textbf {\bibinfo {volume} {27}},\ \bibinfo
  {pages} {745} (\bibinfo {year} {1971})}\BibitemShut {NoStop}%
\bibitem [{\citenamefont {Goldemberg}\ \emph {et~al.}(1963)\citenamefont
  {Goldemberg}, \citenamefont {Torizuka}, \citenamefont {Barber},\ and\
  \citenamefont {Walecka}}]{GOLDEMBERG1963242}%
  \BibitemOpen
  \bibfield  {author} {\bibinfo {author} {\bibfnamefont {J.}~\bibnamefont
  {Goldemberg}}, \bibinfo {author} {\bibfnamefont {Y.}~\bibnamefont
  {Torizuka}}, \bibinfo {author} {\bibfnamefont {W.~C.}\ \bibnamefont
  {Barber}}, \ and\ \bibinfo {author} {\bibfnamefont {J.~D.}\ \bibnamefont
  {Walecka}},\ }\href {\doibase https://doi.org/10.1016/0029-5582(63)90345-6}
  {\bibfield  {journal} {\bibinfo  {journal} {Nucl. Phys.}\ }\textbf {\bibinfo
  {volume} {43}},\ \bibinfo {pages} {242 } (\bibinfo {year}
  {1963})}\BibitemShut {NoStop}%
\bibitem [{\citenamefont {Varlamov}\ \emph {et~al.}(2012)\citenamefont
  {Varlamov}, \citenamefont {Komarov}, \citenamefont {Peskov},\ and\
  \citenamefont {Stepanov}}]{Varmalov2012}%
  \BibitemOpen
  \bibfield  {author} {\bibinfo {author} {\bibfnamefont {V.~V.}\ \bibnamefont
  {Varlamov}}, \bibinfo {author} {\bibfnamefont {S.~Y.}\ \bibnamefont
  {Komarov}}, \bibinfo {author} {\bibfnamefont {N.~N.}\ \bibnamefont {Peskov}},
  \ and\ \bibinfo {author} {\bibfnamefont {M.~E.}\ \bibnamefont {Stepanov}},\
  }\href {\doibase 10.3103/S1062873812040351} {\bibfield  {journal} {\bibinfo
  {journal} {Bull. Russ. Acad. Sci. Phys.}\ }\textbf {\bibinfo {volume} {76}},\
  \bibinfo {pages} {491} (\bibinfo {year} {2012})}\BibitemShut {NoStop}%
\bibitem [{\citenamefont {Ahrens}\ \emph {et~al.}(1975)\citenamefont {Ahrens}
  \emph {et~al.}}]{Ahrens:1975rq}%
  \BibitemOpen
  \bibfield  {author} {\bibinfo {author} {\bibfnamefont {J.}~\bibnamefont
  {Ahrens}} \emph {et~al.},\ }\href {\doibase 10.1016/0375-9474(75)90543-6}
  {\bibfield  {journal} {\bibinfo  {journal} {Nucl. Phys. A}\ }\textbf
  {\bibinfo {volume} {251}},\ \bibinfo {pages} {479} (\bibinfo {year}
  {1975})}\BibitemShut {NoStop}%
\bibitem [{\citenamefont {Gorchtein}(2020)}]{Gorch}%
  \BibitemOpen
  \bibfield  {author} {\bibinfo {author} {\bibfnamefont {M.}~\bibnamefont
  {Gorchtein}},\ }\href@noop {} {}\bibinfo {howpublished} {private
  communication} (\bibinfo {year} {2020})\BibitemShut {NoStop}%
\bibitem [{\citenamefont {Gorchtein}(2021)}]{Gorch2}%
  \BibitemOpen
  \bibfield  {author} {\bibinfo {author} {\bibfnamefont {M.}~\bibnamefont
  {Gorchtein}},\ }\href@noop {} {}\bibinfo {howpublished} {private
  communication} (\bibinfo {year} {2021})\BibitemShut {NoStop}%
\bibitem [{\citenamefont {Koshchii}\ \emph {et~al.}(2021)\citenamefont
  {Koshchii}, \citenamefont {Gorchtein}, \citenamefont {Roca-Maza},\ and\
  \citenamefont {Spiesberger}}]{Koshchii:2021mqq}%
  \BibitemOpen
  \bibfield  {author} {\bibinfo {author} {\bibfnamefont {O.}~\bibnamefont
  {Koshchii}}, \bibinfo {author} {\bibfnamefont {M.}~\bibnamefont {Gorchtein}},
  \bibinfo {author} {\bibfnamefont {X.}~\bibnamefont {Roca-Maza}}, \ and\
  \bibinfo {author} {\bibfnamefont {H.}~\bibnamefont {Spiesberger}},\
  }\href@noop {} {\  (\bibinfo {year} {2021})},\ \Eprint
  {http://arxiv.org/abs/2102.11809} {arXiv:2102.11809 [nucl-th]} \BibitemShut
  {NoStop}%
\end{thebibliography}%

\end{document}